\documentclass[aps,prd,reprint,nofootinbib,superscriptaddress]{revtex4-2}

\usepackage{amsmath,amssymb,bm,stmaryrd}
\usepackage{mathrsfs}

\newcommand{\Lie}{\mathcal{L}}
\newcommand{\eom}{\approx}
\newcommand{\dd}{\mathrm{d}}
\newcommand{\half}{\tfrac12}
\newcommand{\Ds}{\mathcal{D}^{\ast}}

\usepackage[dvipsnames]{xcolor}
\usepackage{hyperref}
\hypersetup{colorlinks=true, linkcolor=Blue, citecolor=Blue, filecolor=Blue, urlcolor=Blue}
\usepackage[capitalize,nameinlink]{cleveref}
\crefname{equation}{Eq.}{Eqs.}
\Crefname{equation}{Equation}{Equations}
\crefname{figure}{Fig.}{Figs.}
\Crefname{figure}{Figure}{Figures}
\crefname{section}{Sec.}{Secs.}
\Crefname{section}{Section}{Sections}
\crefname{table}{Table}{Tables}
\Crefname{table}{Table}{Tables}
\crefname{appendix}{Appendix}{Appendices}
\Crefname{appendix}{Appendix}{Appendices}
\creflabelformat{equation}{#2(#1)#3}

\begin{document}

\title{Internal translation gauging and local gravitational energy\\
in Weyl gauge theory}

\author{Michael \surname{Hobson}}
\email{mph@mrao.cam.ac.uk}
\affiliation{Astrophysics Group, Cavendish Laboratory, J.J. Thomson Avenue, Cambridge, CB3 0HE, UK}

\author{Will Barker}
\email{barker@fzu.cz}
\affiliation{Institute of Physics of the Czech Academy of Sciences, Na Slovance 1999/2, 182 00 Prague 8, Czechia}

\author{Anthony \surname{Lasenby}}
\email{a.n.lasenby@mrao.cam.ac.uk}
\affiliation{Astrophysics Group, Cavendish Laboratory, J.J. Thomson Avenue, Cambridge, CB3 0HE, UK}
\affiliation{Kavli Institute for Cosmology, Madingley Road, Cambridge, CB3 0HA, UK}

\date{\today}

\begin{abstract}
We investigate the local gravitational source and charges obtained by introducing translations as an internal St\"uckelberg redundancy in Weyl gauge theory. The internal-$T(4)$-invariant dressed coframe identifies the source of the translational potential with the existing coframe variational derivative, without introducing new dynamics. We derive an excitation/source equation and balance law whose structure is independent of the constitutive parameters of the parity-even quadratic torsion family. The gravitational self-current, defined relative to a specified gravitational/matter split and boundary representative, gives a density invariant under the internal gauge factors once its frame index is contracted with the coframe; this invariance neither establishes ordinary conservation nor identifies the density with the external translation Noether current. A surface charge follows by contraction with a covariantly constant internal direction, and we separately derive the three-sector Noether superpotential of generalized reducibility, with the asymptotic conditions for conservation of its flux. At the teleparallel equivalent of general relativity (TEGR) the excitation reduces to the teleparallel superpotential and the covariant coframe--connection pair gives the Schwarzschild ADM energy; flat cosmology and transverse-traceless waves give further checks. Away from TEGR, an exact solution of New General Relativity gives a charge whose fractional difference from the Kepler mass is half the deviation of the post-Newtonian spatial-curvature parameter from unity, at the adopted normalisation; independent Hamiltonian charge matching remains open. Finally, the integrable Hamiltonian charge is invariant under covariant Lagrangian boundary terms at fixed reference normalisation and presymplectic representative.
\end{abstract}

\maketitle

\section{Introduction}
\label{sec:intro}

The localisation of gravitational energy depends on the symmetry and the
class of local expressions under consideration. In the preceding papers
of this series, we applied the Bessel-Hagen construction to Fierz-Pauli
and linearised Weyl-squared gravity~\cite{HobsonBarkerLasenbyFP,HobsonBarkerLasenbyWS}.
In both cases the construction gives a gauge-invariant class of currents,
but no preferred strictly invariant local energy-momentum tensor. The
first theory contains too few derivatives to form the required invariant
curvature bilinear, whereas in the second a four-dimensional identity
reduces the allowed rank-two Weyl bilinears to pure traces. These examples
were placed in a common framework in Ref.~\cite{HobsonBarkerLasenbyEMT}:
for Fierz-Pauli and dynamically regular pure curvature-squared metric
theories in four dimensions, homogeneity in total derivative number
allows projection by field degree and derivative number to exclude every
strictly invariant representative of the nontrivial translation class
under the stated locality, polynomiality, finite-order,
Poincar\'e-covariance and regularity assumptions. Constant translations
independently determine linear reducibility charges. Reducibility does
not imply the exclusion, and the theorem does not cover arbitrary gauge
theories or mixed derivative-order actions.

Those analyses concern translations realised externally, whose
localisation gives general coordinate transformations. The tetrad then
provides a soldering map, rather than a Yang-Mills connection for an
internal translation group. Electromagnetism supplies the comparison:
its internal gauge symmetry and first-derivative invariant field strength
allow the Bessel-Hagen construction to yield the Maxwell tensor. It is
therefore useful to ask what changes when translations are introduced
internally in a gravitational gauge theory with independent first-order
field strengths.

We adopt the interpretation of gauge theories of gravity as field theories
on a fixed Minkowski background, with the tetrad and the other gauge
fields treated on the same footing as the fields of the other
interactions~\cite{HobsonLasenbyBarker2024,CGTpaper,eWGTpaper}. This
viewpoint applies also to Poincar\'e gauge theory
(PGT)~\cite{Kibble1961,BlagojevicHehl2013}. The tetrad ${h_a}^\mu$ and
its inverse ${b^a}_\mu$ relate coordinate and local Lorentz indices.
Following the gauge-gravity and teleparallel
literature~\cite{HehlMcCreaMielkeNeeman1995,ItinHehlObukhov2017}, we
call ${b^a}_\mu$ the coframe: the one-forms ${b^a}_\mu\dd x^\mu$ are
dual to ${h_a}^\mu\partial_\mu$, and the coframe is itself a dynamical
variable. Its decomposition into a translational potential and the
covariant derivative of a compensating field introduces an internal
$T(4)$ redundancy of St\"uckelberg type. Related constructions using
parameterization, or clock, scalars have been considered by Koivisto and
Z\l{}o\'snik~\cite{KoivistoZlosnik2022,KoivistoZlosnik2024}. Their
pregeometric setting has a coframe composed entirely of such derivatives;
here an independent coframe remains, and we examine its source and charge
structure.

We work in Weyl gauge theory (WGT), although the internal translation
construction also applies to PGT on removing the dilation sector. Retaining
that sector allows us to derive the balance law and generalized-reducibility
superpotential for the full Weyl multiplet, with Lorentz curvature and
dilational field strength both present. It also provides the appropriate
starting point for a later treatment of extended Weyl gauge theory
(eWGT), whose modification concerns the dilation and torsion-trace sector.
The dilation field supplies a further comparison with electromagnetism:
its abelian internal symmetry has a first-order field strength
$H_{\mu\nu}$, and the associated Maxwell-type mixed coordinate stress
density has Weyl weight zero, whereas the undensitised tensor has weight
$-4$~\cite{HobsonLasenbyBarker2024}. A full Bessel-Hagen analysis of
that sector, suggested in Ref.~\cite{HobsonBarkerLasenbyWS}, is not
undertaken here.

The internal $T(4)$ leaves the dressed coframe and its field strengths
inert, so their strict invariance is kinematical. The redundancy produces
neither new dynamics nor a Noether current of its own. Its translational
potential instead has the coframe variational derivative as its source.
We examine what this interpretation establishes about that existing
source, working nonlinearly so that its full excitation/source equation,
balance law and charge constructions can be retained. The linearised
comparison is discussed in \cref{sec:internal}; it does not identify
the internal redundancy with the external metric gauge symmetry at any
order. In particular, the present construction neither supplies nor
excludes a strict representative of the external translation class of the
independent-connection theory. None of the results below depends on the
metric no-go theorem, which serves as motivation and as a comparison in
\cref{app:torsionfree}.

The variational structure on which we rely is familiar from Poincar\'e
and metric-affine gauge theories~\cite{HehlMcCreaMielkeNeeman1995,Obukhov2018,BlagojevicHehl2013}.
The excitation conjugate to a field strength is, up to normalisation,
the derivative of the Lagrangian with respect to that field strength;
the field equation equates its divergence to a source. This is analogous
to the electromagnetic excitation $(\mathbf D,\mathbf H)$ conjugate
to $(\mathbf E,\mathbf B)$. In premetric teleparallel gravity the
relation between torsion and excitation is specified by a constitutive
law~\cite{ItinHehlObukhov2017}. Here we use this structure to define a
gravitational self-current relative to a specified gravitational/matter
split and boundary representative. The total coframe Euler derivative
vanishes on shell, whereas the total source ${J_a}^\mu$ in the excitation
equation need not. Contracting the frame index of the self-current with
the coframe gives an internally invariant density, but does not by itself
establish its ordinary conservation or select a preferred representative.

The paper proceeds as follows. \Cref{sec:internal} introduces the
internal gauging and distinguishes branch constraints, gauge choices
and truncations. \Cref{sec:theorem} derives the excitation/source
equation and its general balance law, and identifies the soldered density.
\Cref{sec:constitutive,sec:scalar} give the explicit
quadratic torsion family and a scalar matter source. In
\cref{sec:charge} we develop two separate sufficient charge
constructions: contraction with a covariantly constant internal direction,
and generalized reducibility of the full diffeomorphism--Lorentz--dilation
configuration, with a three-sector Noether superpotential for a
scale-invariant quadratic class. Neither condition is proved necessary,
and neither construction is a restriction of the other.

\Cref{sec:tegr} calibrates the charge at the TEGR point against
the Schwarzschild ADM energy and checks the spatially flat cosmology and
transverse-traceless wave. \Cref{sec:discriminator} examines the
constitutive dependence: the wave and cosmology determine
$2\alpha+\beta$ and $2\alpha+\beta+3\gamma$, while the Schwarzschild
charge is unchanged by an axial deformation. A different exact solution
in New General Relativity gives $Q/M=(1+\gamma_{\rm PPN})/2$ with the
adopted normalisation, although independent charge matching for that
model remains open. \Cref{sec:admissible} states the physical
admissibility conditions and proves stability of the reference-normalised
integrable Hamiltonian charge under covariant Lagrangian boundary terms.
We discuss the interpretation and remaining questions in
\cref{sec:discussion}. The torsion-free WGT branch is treated in
\cref{app:torsionfree}.

We use the metric signature $(+\,-\,-\,-)$ and the conventions of
Refs.~\cite{HobsonLasenbyBarker2024,eWGTpaper,CGTpaper}: the tetrad ${h_a}^\mu$ carries
Weyl weight $w=-1$, the determinant $h=\det({h_a}^\mu)$ and the \emph{scalar Lagrangian}
$L$ have weight $w=-4$, and $h^{-1}=\sqrt{-g}$; the object varied below is the
\emph{coordinate Lagrangian density} $\mathcal{L}=h^{-1}L=eL$, of Weyl weight zero in the
scale-invariant formulation. Latin indices are local Lorentz indices
and Greek indices coordinate indices, related by ${h_a}^\mu$ and the coframe ${b^a}_\mu$.
Index conversion by ${h_a}^\mu$ and ${b^a}_\mu$, and the movement of Greek indices with the
composite metric $g_{\mu\nu}\equiv\eta_{ab}{b^a}_\mu{b^b}_\nu$, are used throughout as
definitions of composite fields; in the Minkowski gauge-field reading adopted here they
supply the components of the associated objects in the sense of Sec.~II\,C of
Ref.~\cite{eWGTpaper}, and every invariant employed below is insensitive to that reading.
Following the kernel-font convention of Ref.~\cite{HobsonLasenbyBarker2024}, a quantity
whose indices are all Latin, or a contraction over Latin indices, takes a calligraphic
kernel (${\mathcal{T}^{\ast a}}_{bc}$, $\mathcal{H}_{ab}$), while Greek or mixed index
placement keeps the Roman kernel (${T^{\ast a}}_{\mu\nu}$, $H_{\mu\nu}$); the fraktur
$\mathfrak{H}$ is reserved for the translational excitation introduced in
\cref{sec:theorem}.
Densities and signs are fixed once here. We write $e=\det({b^a}_\mu)$ and
$h=e^{-1}=\det({h_a}^\mu)$, so that a coordinate density of weight one is $e$ times an
ordinary tensor, and frame and coordinate indices are converted with ${b^a}_\mu$ and
${h_a}^\mu$ throughout. The coframe sources below are variational derivatives of a
Lagrangian \emph{density}, and they therefore carry the sign \emph{opposite} to the
corresponding stress tensor: differentiating the scalar action of \cref{sec:scalar}
gives \cref{eq:scalarsource}, which reads ${(t_{\rm m})_a}^\mu=-e\,{h_b}^\mu{T^b}_a$ with
${T^\mu}_\nu=\Ds{}^{\mu}\phi\,\Ds_\nu\phi-\delta^\mu_\nu L_{\rm m}$ the familiar stress
tensor. Consistently, the stress tensor associated with any coframe source ${X_a}^\mu$ is
${T^b}_a=-e^{-1}{b^b}_\mu{X_a}^\mu$, so a \emph{positive} energy density corresponds to a
\emph{negative} time component of the source. The cancellations recorded below, between a
positive matter density and a negative gravitational one, are read with that convention.

Symmetrisation and antisymmetrisation carry unit weight. We write $\eom$ for weak
(on-shell) equality.

\section{Internal gauging of the Weyl group}
\label{sec:internal}

In the usual formulation of gravitational gauge
theories~\cite{Kibble1961,eWGTpaper,CGTpaper,HobsonLasenbyBarker2024},
local translations are realised externally. The coframe also permits an
internal construction on the flat background, in which the homogeneous
Weyl factors act on the fibre and $T(4)$ acts as a St\"uckelberg
redundancy. The latter does not act on matter as an additional Yang-Mills
factor.

Regard local translations as an internal group $T(4)$ acting on the tangent-space (fibre)
coordinates $\xi^a$~\cite{AldrovandiPereira2013,Kibble1961}, with translational potential
${C^a}_\mu$ (kept distinct from the dilational $B_\mu$), Lorentz connection ${A^{ab}}_\mu$,
dilation connection $B_\mu$, and abstract generators $P_a,J_{ab},\Delta$ obeying%
\footnote{These are the \emph{total} generators, which also act on the fibre coordinate
$\xi^a$; on a matter field $J_{ab}$ acts as the intrinsic Lorentz generator $\Sigma_{ab}$
and $\Delta$ as multiplication by the Weyl weight $w$.}
$[J_{ab},P_c]=\eta_{bc}P_a-\eta_{ac}P_b$, $[\Delta,P_a]=P_a$, $[P_a,P_b]=0$. The
translation-invariant, or \emph{dressed}, coframe is built with a compensator
(St\"uckelberg) field $\xi^a$,
\begin{equation}
 {b^a}_\mu=D^{\ast}_\mu\xi^a+{C^a}_\mu ,\qquad
 D^{\ast}_\mu\xi^a=\partial_\mu\xi^a+{A^a}_{b\mu}\xi^b+B_\mu\xi^a ,
\label{eq:t4potential}
\end{equation}
$D^{\ast}_\mu$ being the homogeneous (Lorentz-and-dilation) covariant derivative. Under an
internal translation $\delta_\epsilon\xi^a=\epsilon^a$,
$\delta_\epsilon{C^a}_\mu=-D^{\ast}_\mu\epsilon^a$, with ${A^{ab}}_\mu$ and $B_\mu$ inert,
the compensating shift leaves $\delta_\epsilon{b^a}_\mu=0$, so the coframe, and every field
strength built covariantly from it, is invariant under the internal translations. They act
only in the St\"uckelberg sector $(\xi^a,{C^a}_\mu)$: the coframe and all pulled-back
matter fields are inert, so $P_a$ carries no nontrivial representation on physical matter,
and the symmetry is a redundancy rather than a matter-acting gauge factor.

\subsection{The translational field strength}

The translational field strength is the covariant curl of the coframe,
\begin{equation}
 {T^{\ast a}}_{\mu\nu}=2D^{\ast}_{[\mu}{b^a}_{\nu]}
 =2\partial_{[\mu}{b^a}_{\nu]}+2{A^a}_{b[\mu}{b^{|b|}}_{\nu]}+2B_{[\mu}{b^a}_{\nu]}.
\label{eq:covtorsion}
\end{equation}
Written through the compensator it is the \emph{completed} torsion
\begin{equation}
 {T^{\ast a}}_{\mu\nu}={F^a}_{\mu\nu}
 +{R^{a}}_{b\mu\nu}\,\xi^b+H_{\mu\nu}\,\xi^a ,
\label{eq:completedtorsion}
\end{equation}
where ${F^a}_{\mu\nu}=2D^{\ast}_{[\mu}{C^a}_{\nu]}$ is the curl of the
translational connection, and ${R^{ab}}_{\mu\nu}$ and $H_{\mu\nu}$
are the Lorentz and dilation field strengths. The compensator terms
$R\xi$ and $H\xi$ ensure translation invariance when the homogeneous
curvatures are nonzero. Thus internal translation gauging imposes no
flatness constraint: $T^{\ast}$, $R$ and $H$ may all be nonzero, as in
ordinary WGT. We retain the geometric names torsion and curvature for
these field strengths, while adopting their field-theoretic interpretation
on Minkowski spacetime.

\emph{Relation to the linearised setting.} On the Minkowski background,
${\bar b^a}{}_\mu=\partial_\mu\bar\xi^a+{\bar C^a}{}_\mu=\delta^a_\mu$,
so invertibility prevents both $\bar\xi^a$ and ${\bar C^a}_\mu$ from
being counted as first-order perturbations. In the unitary background
gauge $\bar\xi^a=0$, ${\bar C^a}_\mu=\delta^a_\mu$, the compensator
terms in~\cref{eq:completedtorsion} are second order, and the completed
torsion agrees with the raw curl to first order. The condition
$\delta_\epsilon{C^a}_\mu=-D^{\ast}_\mu\epsilon^a=0$ then selects
constant parameters. Such parameters do not preserve the complete
St\"uckelberg pair, however, since $\delta_\epsilon\xi^a=\epsilon^a$;
the pair has no nonzero reducibility parameter.

These agreements depend on the background gauge and concern only the
linearised kinematics. The internal transformation leaves the coframe
perturbation inert, whereas the external metric gauge symmetry acts on
it. Their symmetry actions and Noether-current classes therefore remain
distinct. The nonlinear analysis below determines the source equation and
charge structure without inferring an identification from this
linearised comparison.

The three connections assemble into a single Yang-Mills operator,
\begin{equation}
 \mathsf{D}^{\ast}_\mu=\partial_\mu+{C^a}_\mu P_a+\half{A^{ab}}_\mu J_{ab}+B_\mu\Delta ,
\label{eq:fullint}
\end{equation}
whose curvature is
\begin{equation}
 [\mathsf{D}^{\ast}_\mu,\mathsf{D}^{\ast}_\nu]={F^a}_{\mu\nu}P_a
 +\half{R^{ab}}_{\mu\nu}J_{ab}+H_{\mu\nu}\Delta .
\label{eq:fullintcomm}
\end{equation}
Its translational component is the \emph{raw} curl ${F^a}_{\mu\nu}$, itself not
invariant under internal translations when $R$ or $H$ is nonzero. The invariant,
soldered field strength is instead the covariant curl \labelcref{eq:covtorsion}; the
curvature-dependent compensator terms in \cref{eq:completedtorsion} are precisely why
neither $R$ nor $H$ need vanish.%
\footnote{The same completion is encoded by a dressed (nonlinear-realisation) connection
$\Gamma_\xi=g^{-1}(\dd+\Gamma)g$ with $g=e^{\xi^aP_a}$. Because $[J_{ab},P_c]$ and
$[\Delta,P_c]$ close on the abelian $P_a$, the conjugation shifts only the translational
component,
$\Gamma_\xi=({b^a}_\mu P_a+\half{A^{ab}}_\mu J_{ab}+B_\mu\Delta)\,\dd x^\mu$,
with the homogeneous parts unchanged; its translational curvature is therefore
${T^{\ast a}}_{\mu\nu}$ directly. On matter fields, on which $P_a$ acts trivially,
$\Gamma_\xi$ thus acts as the homogeneous derivative $D^{\ast}_\mu$, and contraction with
the inverse invariant coframe ${h_a}^\mu$ gives the frame derivative $\Ds_a$. We do not
develop the formalism further.}
The invariant coframe supplies the soldering map between fibre and base: on pulled-back
matter fields the corresponding dressed connection reproduces the standard WGT frame
derivative $\Ds_a$. The internal reading thus reproduces the same coframe coupling as the
external one after an appropriate gauge choice, but it does not identify the two
symmetries, an internal $T(4)$ with finite-dimensional structure group versus the spacetime
diffeomorphisms, or their Noether-current classes.

\subsection{The internal translation source}

Suppose the compensator $\xi^a$ and the translational potential ${C^a}_\mu$ enter the
action only through the coframe ${b^a}_\mu$ and its covariant field strengths. Since
${C^a}_\mu$ then enters only through ${b^a}_\mu$, with
$\delta{b^b}_\nu/\delta{C^a}_\mu=\delta^b_a\delta^\mu_\nu$, the gauge source conjugate to
${C^a}_\mu$ is the coframe energy-momentum current
\begin{equation}
 {t_a}^\mu\equiv\frac{\delta\mathcal{L}}{\delta{C^a}_\mu}
 =\frac{\delta\mathcal{L}}{\delta{b^a}_\mu}=-{h_b}^\mu{\tau^b}_a ,
\label{eq:t4current}
\end{equation}
where ${\tau^a}_\mu=\delta\mathcal{L}/\delta{h_a}^\mu$ is the variational energy-momentum
of Ref.~\cite{HobsonLasenbyBarker2024} and the last equality follows from
$\delta{h_a}^\mu=-{h_a}^\nu{h_b}^\mu\delta{b^b}_\nu$. The internal-translation source is
thus not a new object but the coframe variational derivative already supplied by the
manifestly covariant variational principle, read in the St\"uckelberg variables.

The meaning of this derivative depends on which part of the action is
varied. For the total action, ${t_a}^\mu\eom0$ is the coframe field
equation. The matter variation
$\delta\mathcal{L}_{\mathrm m}/\delta{C^a}_\mu$, by contrast, is the
analogue of a Yang-Mills matter source and is generally nonzero. A
gravitational self-current assigned to a subsystem requires a specified
split of the action and retains improvement dependence. Each is strictly
invariant under the internal $T(4)$ because the coframe is inert. Under
local Lorentz transformations, dilations and coordinate transformations,
the frame-indexed quantities transform homogeneously according to their
index type and Weyl weight~\cite{HobsonLasenbyBarker2024}.

The independent variations of $\xi^a$ and ${C^a}_\mu$ give the internal-translation Noether
identity in coordinate-density form,
\begin{equation}
 \frac{\delta\mathcal{L}}{\delta\xi^a}
 +\partial_\mu{t_a}^\mu-{A^b}_{a\mu}{t_b}^\mu-B_\mu{t_a}^\mu\equiv0 ,
\label{eq:t4noether}
\end{equation}
the connection terms carrying the dual Lorentz and dilational weights of ${t_a}^\mu$.
Because \cref{eq:t4noether} holds identically, the compensator equation
$\delta\mathcal{L}/\delta\xi^a=0$ is not an independent condition: it is the covariant
divergence of the coframe equation. This is Noether's second theorem for the internal
factor, and it is the precise sense in which the redundancy introduces no new dynamics.

The source ${t_a}^\mu$ is conjugate to an internal fibre-translation
potential. Soldering relates its frame index to a spacetime index, but
does not make it the Noether current of rigid translations of the base
coordinates. It is thus a different variational object from the external
translation current considered in Ref.~\cite{HobsonBarkerLasenbyEMT},
and is defined whether or not $R$ and $H$ vanish.

\subsection{The status of the simplifying conditions}

\Cref{tab:status} distinguishes the conditions often imposed on
these fields. A branch constraint restricts the configurations
covariantly; a gauge choice selects a representative on a suitable branch;
and a truncation changes the retained fields or gauge group. None of the
listed flatness or torsion conditions is required to introduce the
internal translation redundancy.

The internal and external descriptions should not be counted as two
independent dynamical gaugings of the same freedom. The St\"uckelberg
construction adds no dynamics and retains coordinate covariance. In the
fixed-background reading, general coordinate transformations express the
covariance of the field theory, whereas rigid Minkowski translations are
a global symmetry with an ordinary Noether current. These must also be
distinguished from the soldered translational component of a spacetime
transformation introduced in \cref{sec:multiplet}. Whether the
external translation class of this independent-connection theory admits
a strictly invariant representative is not settled by the metric theorem
of Ref.~\cite{HobsonBarkerLasenbyEMT} and remains open here.

\begin{table*}
\caption{\label{tab:status}%
Logical status of the conditions arising in the internal gauging of the Weyl group. None
is required merely to gauge translations internally, and the generic case (final row) is
fully allowed: the strictly invariant translational field strength is the
compensator-completed torsion~\labelcref{eq:completedtorsion}, generically nonzero together
with $R$ and $H$.}
\footnotesize
\begin{ruledtabular}
\begin{tabular}{lll}
Condition & Logical status & Consequence\\
\hline
${R^{ab}}_{\mu\nu}=0$ & teleparallel branch constraint & Lorentz connection locally inertial\\
${A^{ab}}_\mu=0$ & local gauge choice (if $R=0$) & Weitzenb\"ock gauge\\
$H_{\mu\nu}=0$ & scale-flat branch constraint & Weyl connection locally pure gauge\\
$B_\mu=0$ & local gauge choice (if $H=0$) & scale connection absent\\
remove $B_\mu$ & truncation / change of group & WGT $\to$ PGT\\
${T^{\ast a}}_{\mu\nu}=0$ & branch constraint (or connection eqn) & curvature-based reduced form\\
$\xi^a=0$ & internal-translation gauge & ${b^a}_\mu={C^a}_\mu$\\
${T^{\ast a}}_{\mu\nu}={\Omega^a}_{\mu\nu}$ & special representative & algebraic condition \labelcref{eq:curlcondition}; $R=H=0$ + local gauges suffices\\
$\{T^{\ast},R,H\}\neq0$ & generic internal WGT & full internally gauged theory\\
\end{tabular}
\end{ruledtabular}
\end{table*}

\emph{Torsion-free reduction.} Imposing ${T^{\ast a}}_{\mu\nu}=0$ is a condition covariant
under the local rotations, dilations and coordinate transformations and strictly
invariant under the internal $T(4)$; it does not remove the internal $T(4)$, which survives as the
St\"uckelberg redundancy $\delta_\epsilon\xi^a=\epsilon^a$ (unitary gauge $\xi^a=0$,
${b^a}_\mu={C^a}_\mu$). Its Bianchi identity constrains only the combination
${R^{a}}_{b[\mu\nu}{b^b}_{\rho]}+H_{[\mu\nu}{b^a}_{\rho]}$, not $R$ or
$H$ separately. In WGT this torsion-free condition defines the ``reduced''
formulation~\cite{eWGTpaper,HobsonLasenbyBarker2024} and solves the
rotational connection explicitly, as the Weyl connection $A[b,B]$;
\cref{app:torsionfree} records the resulting curvature decomposition and shows
that its totally trace-free curvature is the metric Weyl tensor, so that a conformal
curvature term of the action reduces on this branch to its metric counterpart; the
\emph{isolated reduced metric theory} then lies in the domain of the metric
theorem~\cite{HobsonBarkerLasenbyWS,HobsonBarkerLasenbyEMT}, which is a statement about
that isolated theory and not about the current classes of the full reduced system
(\cref{app:torsionfree}). In WGT the dynamics on
this branch is thus curvature-based and the independent first-order translational
field-strength sector is removed: the internal variational source ${t_a}^\mu$ may still
be defined, but is then determined by the reduced curvature-based action rather than by a
nonzero translational field strength. In eWGT the corresponding reduction is instead
imposed on the daggered torsion~\cite{eWGTpaper}; we make no claim about that case here,
and it is treated separately.

\emph{Teleparallelism.} Imposing instead ${R^{ab}}_{\mu\nu}=0$ makes the
Lorentz connection inertial, a pure-gauge
${A^{ab}}_\mu={\Lambda^a}_c\partial_\mu{(\Lambda^{-1})}^{cb}$; this does not remove
the local-Lorentz symmetry, the connection being retained so that covariance stays
manifest~\cite{KrssakSaridakis2016}, but only freezes the curvature. The translational
field strength is retained and first order, so the mechanism permitting a local internal
current remains available, the gravitational dynamics being carried by torsion rather than
curvature.

\emph{Abelian-curl representative.} The completed torsion collapses to the bare object of
anholonomy,
\begin{equation}
 {\Omega^a}_{\mu\nu}=2\partial_{[\mu}{b^a}_{\nu]} ,
\label{eq:torsionfs}
\end{equation}
exactly when the homogeneous connections drop out of \cref{eq:covtorsion},
\begin{equation}
 {A^a}_{b[\mu}{b^{|b|}}_{\nu]}+B_{[\mu}{b^a}_{\nu]}=0 ,
\label{eq:curlcondition}
\end{equation}
that is, $A^a{}_b\wedge b^b+B\wedge b^a=0$. A sufficient realisation is
a Lorentz- and scale-flat branch, with gauges in which both homogeneous
connections vanish. The torsion is then an abelian curl, giving the
closest comparison with electromagnetism. This is only a special
representative: \cref{eq:curlcondition} is an algebraic condition
on connection--coframe contractions and can also hold for configurations
with nontrivial homogeneous fields. It must not be interpreted as a
flatness condition. In particular, the torsion-free branch in
\cref{app:torsionfree} determines the rotational connection
without requiring it to be flat. Removing ${A^{ab}}_\mu$ and $B_\mu$
from the theory, rather than choosing gauges on a flat branch, removes
the associated local symmetries; under a subsequent local change of
frame the bare $\Omega$ then transforms inhomogeneously.

\section{The excitation/source theorem}
\label{sec:theorem}

\subsection{The translations-only torsional sector}

We first consider a translations-only theory with
${b^a}_\mu=\partial_\mu\xi^a+{C^a}_\mu$ and field
strength~\labelcref{eq:torsionfs}. Let its first-order Lagrangian be
\begin{equation}
 \mathcal{L}=\mathcal{L}_{\rm g}({b^a}_\mu,{\Omega^a}_{\mu\nu})
 +\mathcal{L}_{\rm m}({b^a}_\mu,\psi,\partial\psi) ,
\label{eq:transaction}
\end{equation}
in which the matter sector does not couple to the torsion, so that the whole $\Omega$
dependence sits in $\mathcal{L}_{\rm g}$, and define the antisymmetric translational
excitation
\begin{equation}
 {\mathfrak{H}_a}{}^{\mu\nu}\equiv 2\,\frac{\partial\mathcal{L}_{\rm g}}
 {\partial{\Omega^a}_{\mu\nu}}=-{\mathfrak{H}_a}{}^{\nu\mu} .
\label{eq:excitation}
\end{equation}
Since $\delta{\Omega^a}_{\mu\nu}=2\partial_{[\mu}\delta{b^a}_{\nu]}$, integrating by parts
gives
\begin{equation}
 \frac{\delta\mathcal{L}_{\rm g}}{\delta{b^a}_\mu}
 =\left.\frac{\partial\mathcal{L}_{\rm g}}{\partial{b^a}_\mu}\right|_{\Omega}
 -\partial_\nu{\mathfrak{H}_a}{}^{\nu\mu} .
\label{eq:legendre}
\end{equation}
Writing the gravitational translational self-current and the matter source as
\begin{equation}
 {(j_{\rm g})_a}^\mu\equiv
 \left.\frac{\partial\mathcal{L}_{\rm g}}{\partial{b^a}_\mu}\right|_{\Omega} ,\qquad
 {(t_{\rm m})_a}^\mu\equiv\frac{\delta\mathcal{L}_{\rm m}}{\delta{b^a}_\mu} ,
\label{eq:jgtm}
\end{equation}
the coframe (equivalently internal-translation) field equation
$\delta\mathcal{L}/\delta{C^a}_\mu=0$ becomes the Yang-Mills-type source equation
\begin{equation}
 \partial_\nu{\mathfrak{H}_a}{}^{\nu\mu}={(j_{\rm g})_a}^\mu+{(t_{\rm m})_a}^\mu .
\label{eq:sourceeq}
\end{equation}
Because the excitation is antisymmetric,
$\partial_\mu\partial_\nu{\mathfrak{H}_a}{}^{\nu\mu}=0$, and the total translational source
is conserved on that field equation,
\begin{equation}
 \partial_\mu\!\left({j_{\rm g}}+{t_{\rm m}}\right)_a{}^\mu=0 .
\label{eq:transcons}
\end{equation}
In this sector $\delta_\epsilon{b^a}_\mu=0$ and $\delta_\epsilon{\Omega^a}_{\mu\nu}=0$, so
the excitation and both currents are local and strictly invariant under the internal
St\"uckelberg $T(4)$.

\medskip
\noindent\textbf{Theorem (internal translation source current).}\ \emph{Let a
translations-only first-order torsional theory have Lagrangian~\labelcref{eq:transaction} with
${\Omega^a}_{\mu\nu}=2\partial_{[\mu}{b^a}_{\nu]}$ and the St\"uckelberg decomposition
${b^a}_\mu=\partial_\mu\xi^a+{C^a}_\mu$. Then the coframe field equation takes the
form~\labelcref{eq:sourceeq}, with the excitation~\labelcref{eq:excitation} and the
currents~\labelcref{eq:jgtm}; all three are strictly invariant under the internal $T(4)$, and
$\partial_\mu({j_{\rm g}}+{t_{\rm m}})_a{}^\mu=0$ on that field equation.}
\medskip

\noindent The result establishes a conserved total source on the coframe
equation in this translations-only theory. The definition of
${(j_{\rm g})_a}^\mu$ at fixed $\Omega$ retains dependence on the
formulation and boundary term; neither full local Lorentz and Weyl
invariance nor a preferred improvement representative follows. Its
identification with the external rigid-translation current is also a
separate question.

\subsection{The general form and the balance law}

For the general first-order theory, take
$\mathcal{L}_{\rm g}=\mathcal{L}_{\rm g}({b^a}_\mu,
{T^{\ast a}}_{\mu\nu},{R^{ab}}_{\mu\nu},H_{\mu\nu})$ and define
the excitation and explicit coframe derivative at fixed field strengths by
\begin{equation}
 {\mathfrak{H}_a}{}^{\mu\nu}\equiv2\,\frac{\partial\mathcal{L}_{\rm g}}
 {\partial{T^{\ast a}}_{\mu\nu}},\qquad
 {(j_{\rm g})_a}^\mu\equiv\left.\frac{\partial\mathcal{L}_{\rm g}}
 {\partial{b^a}_\mu}\right|_{T^{\ast},R,H} ,
\label{eq:genexc}
\end{equation}
extending \cref{eq:excitation,eq:jgtm}, with
${(t_{\rm m})_a}^\mu=\delta\mathcal{L}_{\rm m}/\delta{b^a}_\mu$ the matter coframe source
as before. Since ${T^{\ast a}}_{\mu\nu}=2D^{\ast}_{[\mu}{b^a}_{\nu]}$ gives
$\delta{T^{\ast a}}_{\mu\nu}=2D^{\ast}_{[\mu}\delta{b^a}_{\nu]}$ at fixed
homogeneous connections, one integration by parts yields the coframe equation in
excitation/source form,
\begin{equation}
 (\Ds_\nu+T^{\ast}_\nu)\,{\mathfrak{H}_a}{}^{\nu\mu}
 ={(j_{\rm g})_a}^\mu+{(t_{\rm m})_a}^\mu\equiv{J_a}^\mu ,
\label{eq:gensource}
\end{equation}
the operator $(\Ds_\nu+T^{\ast}_\nu)$ being the same combination that converts a
covariant into an ordinary divergence in the conservation laws of
Ref.~\cite{HobsonLasenbyBarker2024}. Written out on a coordinate density with a lower
frame index of Weyl weight $-1$, it is
\begin{equation}
 (\Ds_\nu+T^{\ast}_\nu)\,{X_a}^{\nu\mu}
 =\partial_\nu{X_a}^{\nu\mu}-{A^b}_{a\nu}{X_b}^{\nu\mu}-B_\nu{X_a}^{\nu\mu} ,
\label{eq:dtexpand}
\end{equation}
so the coordinate-density divergence contains the partial derivative and
the homogeneous connection terms shown explicitly. The notation
$(\Ds_\nu+T^{\ast}_\nu)$ denotes this combination; it does not add a
further torsion trace to the right-hand side.%
\footnote{In exterior-form notation the field equation reads
$D^{\ast}\mathfrak{H}_a=E_a+\Sigma_a$, where
$\mathfrak{H}_a=-\partial V/\partial T^a$,
$E_a=\partial V/\partial\vartheta^a|_{T,R,H}$ and $\Sigma_a$ is the
matter coframe source. The integration by parts is then immediate.
The two-form $\mathfrak{H}_a$ is not numerically identified with the
coordinate density ${\mathfrak{H}_a}{}^{\mu\nu}$; their relation uses
the orientation and normalisation stated after~\cref{eq:multcharge}.
No factor of $e$ occurs there, since the alternating symbol is used.}

Under the internal translations
$\delta_\epsilon{b^a}_\mu=\delta_\epsilon{T^{\ast a}}_{\mu\nu}
=\delta_\epsilon{R^{ab}}_{\mu\nu}=\delta_\epsilon H_{\mu\nu}=0$, whence
\begin{equation}
 \delta_\epsilon{\mathfrak{H}_a}{}^{\mu\nu}=\delta_\epsilon{(j_{\rm g})_a}^\mu
 =\delta_\epsilon{(t_{\rm m})_a}^\mu=\delta_\epsilon{J_a}^\mu=0 .
\label{eq:genstrict}
\end{equation}
\cref{eq:gensource} is the usual coframe field equation of
gravitational gauge theory~\cite{HehlMcCreaMielkeNeeman1995,Obukhov2018,BlagojevicHehl2013};
with flat homogeneous connections it gives the premetric teleparallel
relation for a specified constitutive law~\cite{ItinHehlObukhov2017}.
The internal construction makes explicit that the excitation, self-current,
matter source and total source are each invariant under the introduced
$T(4)$, as stated in~\cref{eq:genstrict}.

The total translational source is not generally separately conserved away from a flat branch.
Write the coframe Euler derivative as
\begin{equation}
 {\mathcal{E}_a}^\mu\equiv{J_a}^\mu-(\Ds_\nu+T^{\ast}_\nu){\mathfrak{H}_a}{}^{\nu\mu} ,
\label{eq:coframeeuler}
\end{equation}
so that \cref{eq:gensource} is ${\mathcal{E}_a}^\mu=0$. Taking a further divergence, the
antisymmetry of the excitation cancels the double derivative and the Ricci identity
converts the remainder into contractions with the homogeneous field strengths. The exact
\emph{off-shell} identity is
\begin{equation}
\begin{split}
 (\Ds_\mu+T^{\ast}_\mu){J_a}^\mu
 ={}&(\Ds_\mu+T^{\ast}_\mu){\mathcal{E}_a}^\mu\\
 &+\half\,{R^b}_{a\mu\nu}{\mathfrak{H}_b}{}^{\mu\nu}
 +\half\,H_{\mu\nu}{\mathfrak{H}_a}{}^{\mu\nu} ,
\end{split}
\label{eq:balanceoff}
\end{equation}
and the balance law
\begin{equation}
 (\Ds_\mu+T^{\ast}_\mu){J_a}^\mu
 =\half\,{R^b}_{a\mu\nu}{\mathfrak{H}_b}{}^{\mu\nu}
 +\half\,H_{\mu\nu}{\mathfrak{H}_a}{}^{\mu\nu}
\label{eq:balance}
\end{equation}
follows on imposing the coframe equation. The off-shell
identity~\labelcref{eq:balanceoff} uses only antisymmetry and the divergence
operator~\labelcref{eq:dtexpand}; it therefore holds for any antisymmetric
excitation, independently of a constitutive law, and contains no residual
torsion term on its right-hand side. Replacing the excitation divergence
by the source requires the field equation. We have checked the identity
symbolically both for an arbitrary antisymmetric excitation and for the
quadratic constitutive family of \cref{sec:constitutive}, with all
three coefficients arbitrary, at generic field configurations.

On a Lorentz- and scale-flat branch, $R=H=0$ makes the right-hand side
of~\cref{eq:balance} vanish. Choosing gauges in which the flat
connections vanish then gives~\cref{eq:sourceeq,eq:transcons}.
For nonzero homogeneous curvatures the source instead obeys a balance
law, analogous to the scalar matter law in \cref{sec:scalar}.
Its curvature terms describe exchange with the homogeneous sectors when
those sectors are dynamical. If the connections are auxiliary,
constrained or eliminated algebraically, the equation is a reduced-system
identity and need not describe physical transfer of energy.

\subsection{The soldered gravitational density}

The frame-indexed self-current transforms covariantly under the homogeneous
gauge factors. Contracting it with the coframe gives the mixed coordinate
density
\begin{equation}
 \widehat{\jmath}_\nu{}^\mu\equiv{b^a}_\nu\,{(j_{\rm g})_a}^\mu ,
\label{eq:soldered}
\end{equation}
which is locally Lorentz invariant, since the coframe and source transform
in dual representations. Both factors are separately invariant under the
internal $T(4)$, so their product is also strictly translation invariant.
In a scale-invariant formulation with a compensator $\chi$ of weight
$-1$, as in \cref{sec:multiplet}, invariance of $\mathcal{L}_{\rm g}$ under
$({b^a}_\mu,{T^{\ast a}}_{\mu\nu},\chi)\to
(e^\sigma{b^a}_\mu,e^\sigma{T^{\ast a}}_{\mu\nu},e^{-\sigma}\chi)$
at fixed $R$ and $H$ gives
${(j_{\rm g})'_a}^\mu=e^{-\sigma}{(j_{\rm g})_a}^\mu$.
Consequently $\widehat{\jmath}_\nu{}^\mu$ has Weyl weight zero and is
also dilation invariant. This last statement applies to the
scale-invariant theory, whereas the TEGR calibration below removes the
scale sector and freezes $\chi^2=\kappa^{-1}$.

The soldered object is a coordinate $(1,1)$ tensor density of weight one;
$e^{-1}\widehat{\jmath}_\nu{}^\mu$ is an ordinary tensor of Weyl weight
$-4$. Soldering uses a field of the theory and requires neither an
externally specified direction in the internal space nor a gauge choice. It does not, however,
imply ordinary conservation of this gravitational density: the balance
law concerns the total frame-indexed source, and differentiating the
coframe introduces an additional term. We return to that distinction in
\cref{sec:discussion}. The definition also retains the specified
action split and boundary representative and does not identify the
density with the external translation Noether current.

\section{Constitutive universality}
\label{sec:constitutive}

The excitation/source equation fixes the form of the field equation,
while a constitutive law specifies the excitation as a function of the
field strength. In a Lagrangian theory that relation follows
from~\cref{eq:genexc}, as in premetric electrodynamics and teleparallel
gravity~\cite{ItinHehlObukhov2017}. We now obtain the excitation and
self-current explicitly for the parity-even quadratic torsion family.
The constitutive relation is linear in torsion, with three independent
coefficients, and the source equation selects none of them.

This family is the direct first-order-field-strength analogue of the
Maxwell action. Its three invariants exhaust the parity-even quadratic
torsion scalars in four dimensions and are equivalent to the squares of
the three irreducible torsion parts~\cite{HehlMcCreaMielkeNeeman1995,Obukhov2018}.
We omit the two parity-odd invariants. The family includes TEGR, needed
for the calibration in \cref{sec:tegr}. Additional dependence on
$R$ and $H$ is allowed by the general source equation, but is not needed
to exhibit this translational constitutive freedom.

In components write $\mathcal{L}_{\rm g}=\tfrac{e}{2\kappa}\,\mathbb{T}$ with
\begin{equation}
 \mathbb{T}=\alpha A_1+\beta A_2+\gamma A_3 ,
\label{eq:family}
\end{equation}
\begin{equation}
 A_1=T_{\lambda\mu\nu}T^{\lambda\mu\nu},\quad
 A_2=T_{\lambda\mu\nu}T^{\mu\lambda\nu},\quad
 A_3=v_\mu v^\mu ,
\label{eq:invariants}
\end{equation}
$v_\mu={T^\lambda}_{\lambda\mu}$ and $e=\det{b^a}_\mu=h^{-1}=\sqrt{-g}$. The teleparallel equivalent of general
relativity is the single point $(\alpha,\beta,\gamma)=(\tfrac14,\tfrac12,-1)$. Two
statements hold for the whole family.

First, since $\mathbb{T}$ depends on the torsion only through the antisymmetric
${T^a}_{\mu\nu}$, the excitation \labelcref{eq:excitation} is antisymmetric for \emph{every}
$(\alpha,\beta,\gamma)$; conservation \labelcref{eq:transcons} therefore holds throughout the
family on the coframe equation in the translations-only sector, independently
of the TEGR choice. In the general theory the corresponding statement is
the balance law~\labelcref{eq:balance}.

Second, the gravitational self-current is the canonical coframe expression of
Refs.~\cite{HehlMcCreaMielkeNeeman1995,Obukhov2018}, and it and the general superpotential
${S_\lambda}^{\mu\nu}=\half\,\partial\mathbb{T}/\partial{T^\lambda}_{\mu\nu}$ are both
fixed explicitly, with no undetermined coefficients beyond the constitutive
$(\alpha,\beta,\gamma)$:
\begin{equation}
 {S_\lambda}^{\mu\nu}=\alpha\,{T_\lambda}^{\mu\nu}
 +\beta\,T^{[\mu}{}_\lambda{}^{\nu]}+\gamma\,\delta_\lambda^{[\mu}v^{\nu]} ,
\label{eq:superpot}
\end{equation}
\begin{equation}
 {(j_{\rm g})_\lambda}^\mu=\frac{e}{2\kappa}\Big(\delta_\lambda^\mu\,\mathbb{T}
 +4\,{T^\rho}_{\lambda\sigma}\,{S_\rho}^{\sigma\mu}\Big) .
\label{eq:hehlform}
\end{equation}
Here $T^{\mu}{}_\lambda{}^{\nu}\equiv g^{\nu\sigma}{T^\mu}_{\lambda\sigma}$
retains the first index raised and the second lowered; lowering the
first index and raising the other two would instead give
${T_\lambda}^{\mu\nu}$. \Cref{eq:hehlform} is the soldered
self-current~\labelcref{eq:soldered}. Both expressions have been checked by
exact symbolic calculation at a generic constant invertible coframe
with arbitrary torsion, separately for $A_1$, $A_2$ and $A_3$. A third
candidate contraction has zero coefficient. Thus the algebraic form is
common to the family, while its value depends on the constitutive
coefficients through $S$.

The compact expression familiar from the teleparallel
literature~\cite{deAndradeGuillenPereira2000},
\begin{equation}
 {K_\lambda}^\mu\equiv{S_\sigma}^{\nu\mu}{T^\sigma}_{\nu\lambda}
 -\tfrac14\delta^\mu_\lambda\mathbb{T} ,
\label{eq:compact}
\end{equation}
is not special to the TEGR point: it is \cref{eq:hehlform} rewritten. Torsion is
antisymmetric in its last two indices, so
${S_\sigma}^{\nu\mu}{T^\sigma}_{\nu\lambda}=-{T^\sigma}_{\lambda\nu}{S_\sigma}^{\nu\mu}$,
whence $-4{K_\lambda}^\mu=\delta^\mu_\lambda\mathbb{T}
+4{T^\sigma}_{\lambda\nu}{S_\sigma}^{\nu\mu}$, which is the bracket of
\cref{eq:hehlform}, and therefore
\begin{equation}
 {(j_{\rm g})_\lambda}^\mu=-\frac{2e}{\kappa}\,{K_\lambda}^\mu .
\label{eq:compactdict}
\end{equation}
This rearrangement uses only the antisymmetry of torsion and holds for
any ${S_\lambda}{}^{\mu\nu}$. The factor $-4$ converts the
normalisation $e/2\kappa$ in~\cref{eq:hehlform} to $-2e/\kappa$
in~\cref{eq:compactdict}; no TEGR-specific constitutive relation is
used. The premetric teleparallel
framework~\cite{ItinHehlObukhov2017} makes this freedom explicit.
TEGR will therefore serve as the point at which the general construction
can be compared with an independently known charge.

\section{A scalar matter source}
\label{sec:scalar}

An explicit matter model illustrates the source appearing
in~\cref{eq:sourceeq,eq:gensource}. Consider a real
scalar of Weyl weight $w=-1$ with Lagrangian
\begin{equation}
 L_{\rm m}=\half\,\Ds_a\phi\,\Ds{}^{a}\phi-\frac{\lambda}{4}\phi^4 ,
\label{eq:scalarL}
\end{equation}
where $\Ds_a\phi={h_a}^\mu(\partial_\mu+wB_\mu)\phi$ and frame indices are raised with
$\eta^{ab}$. The scalar carries no intrinsic spin current, so the
internal-translation matter source \labelcref{eq:t4current} is
\begin{equation}
 {(t_{\rm m})_a}^\mu=-h^{-1}{h_b}^\mu\!\left(\Ds{}^{b}\phi\,\Ds_a\phi
 -\delta^b_a L_{\rm m}\right),
\label{eq:scalarsource}
\end{equation}
\emph{minus} the density form of the familiar stress tensor
$\Ds{}^{\mu}\phi\,\Ds_\nu\phi-\delta^\mu_\nu L_{\rm m}$, as the variational convention
above requires. Because ${b^a}_\mu$
and $\Ds\phi$ are inert under the internal translations, the source is
strictly invariant, $\delta_\epsilon{(t_{\rm m})_a}^\mu=0$; it is nonzero,
a null profile $\phi=f(k\cdot x)$ with $k^2=0$ at $\lambda=0$ in the flat
gauge giving the stress tensor $k^\mu k_\nu[f'(k\cdot x)]^2$; and it is not
ordinarily conserved on a general background. Specialising the covariant
conservation law of Ref.~\cite{HobsonLasenbyBarker2024} to the scalar, for
which the spin current vanishes and the matter field equation holds, gives
the balance law
\begin{equation}
 (\Ds_c+\mathcal{T}^{\ast}_c)(h\,{\tau^c}_d)
 =h\!\left(\zeta^c\mathcal{H}_{cd}-{\tau^c}_b{\mathcal{T}^{\ast b}}_{cd}\right),
\label{eq:scalarbalance}
\end{equation}
with $h\,{\tau^a}_b=\Ds{}^{a}\phi\,\Ds_b\phi-\delta^a_b L_{\rm m}$ and
$\zeta^c=h^{-1}w\,\phi\,\Ds{}^{c}\phi$ the scalar's dilation current, given the same
density character as ${\tau^a}_b$ so that the common factor $h$ in
\cref{eq:scalarbalance} applies uniformly to both: matter
exchanges energy-momentum with the dilation and torsion gauge fields, and
the homogeneous covariant law $\Ds_c(h\,{\tau^c}_d)=0$ in particular holds in the
force-free limit $\mathcal{H}_{ab}=0$, ${\mathcal{T}^{\ast a}}_{bc}=0$, which is
sufficient but not necessary: the force contractions can vanish for particular matter
configurations. The matter source is therefore local and strictly internal-$T(4)$
invariant, but generally satisfies a balance law rather than an ordinary
conservation law. Its rotational source vanishes because the scalar has
no intrinsic spin; its dilational source appears in the exchange term.
The coefficients in~\cref{eq:scalarbalance} follow from the covariant
conservation law of Ref.~\cite{HobsonLasenbyBarker2024}; the density
conventions above fix their use here.

\section{Charge and generalized reducibility}
\label{sec:charge}

The source equation constrains the Lorentz-vector-valued density
${J_a}^\mu$. To obtain a scalar charge, one may contract it with an
internal direction $n^a$. Differentiation then also acts on $n^a$, so
conservation requires additional conditions on the direction and the
configuration. We first derive a sufficient condition using the source
equation, then treat generalized reducibility of the full gauge
configuration as a separate construction.

The parameters and currents involved have different meanings.
The St\"uckelberg parameter $\epsilon^a$ acts only on the redundant
pair; ${t_a}^\mu$ is the coframe variational derivative conjugate to
${C^a}_\mu$; and ${J_a}^\mu=(j_{\rm g})_a{}^\mu+{(t_{\rm m})_a}^\mu$
is the total source in~\cref{eq:gensource}. The direction $n^a$ used
for contraction is independent of these gauge parameters. In the second
construction, $\epsilon_\xi^a=\xi^\mu{b^a}_\mu$ is instead fixed by
a spacetime generator and the coframe. Keeping these definitions distinct
is necessary when comparing the source, symmetry and charge.

\subsection{The charge identity}

Let $n^a$ be an internal translation direction. On the coframe equation,
its contraction with the source can be expressed as a surface term and
a term involving its derivative. In form notation, provided
$n^a\mathfrak{H}_a$ has total Weyl weight zero, the Leibniz rule
$\dd(n^a\mathfrak{H}_a)=D^{\ast}n^a\wedge\mathfrak{H}_a+n^aD^{\ast}\mathfrak{H}_a$ gives
\begin{equation}
 n^aJ_a=\dd(n^a\mathfrak{H}_a)-D^{\ast}n^a\wedge\mathfrak{H}_a .
\label{eq:chargeid}
\end{equation}
In components the same statement, obtained by contracting \cref{eq:gensource} with
$n^a$, reads
\begin{equation}
 n^a{J_a}^\mu=\partial_\nu\!\big(n^a{\mathfrak{H}_a}{}^{\nu\mu}\big)
 -\big(D^{\ast}_\nu n^a\big){\mathfrak{H}_a}{}^{\nu\mu} .
\label{eq:chargeidcomp}
\end{equation}
The additional term contains
$\big(D^{\ast}_\nu n^a\big){\mathfrak{H}_a}{}^{\nu\mu}$.
Covariant constancy $D^{\ast}_\nu n^a=0$ is sufficient to remove it,
but not necessary: the contraction may vanish for a nonparallel direction,
and further cancellations may occur on taking the divergence of the
scalar current. Under the sufficient condition, the charge is
\begin{equation}
 Q[n]=\oint_{\partial\Sigma}n^a{\mathfrak{H}_a}{}^{i0}\,\dd S_i ,
\label{eq:charge}
\end{equation}
which is the form evaluated in \cref{sec:tegr}. Without that condition the exact
statement on the coframe equation is
\begin{equation}
 \int_\Sigma n^a{J_a}^0\,\dd^3x
 =\oint_{\partial\Sigma}n^a{\mathfrak{H}_a}{}^{i0}\,\dd S_i
 -\int_\Sigma\big(D^{\ast}_i n^a\big){\mathfrak{H}_a}{}^{i0}\,\dd^3x ,
\label{eq:bulkboundary}
\end{equation}
so equality with the surface integral requires the last integral to
vanish or to be explicitly subtracted. Pointwise vanishing is sufficient,
but not necessary. Asymptotic parallelism $D^{\ast}_\nu n^a\to0$
controls only the boundary behaviour and does not remove this interior
contribution. Accordingly, conditions for the surface flux and for its
equality to a volume integral must be specified separately.

The order of the excitation indices fixes the sign. Setting $\mu=0$
in~\cref{eq:chargeidcomp} leaves
$\partial_i(n^a{\mathfrak{H}_a}{}^{i0})$, since antisymmetry removes
the $\nu=0$ term. The outward charge flux therefore uses
${\mathfrak{H}_a}{}^{i0}=-{\mathfrak{H}_a}{}^{0i}$, as in the
Schwarzschild calculation below.

The integrability condition for
$D^{\ast}_\mu n^a=0$ is the Ricci identity for the homogeneous connection,
\begin{equation}
 2D^{\ast}_{[\mu}D^{\ast}_{\nu]}n^a
 =\big({R^a}_{b\mu\nu}+w_n\,\delta^a_b\,H_{\mu\nu}\big)n^b=0 ,
\label{eq:integrability}
\end{equation}
which requires the holonomy of the homogeneous connection to preserve
$n^a$. The connection here is the independent WGT homogeneous connection,
not the Levi-Civita connection. A Lorentz- and scale-flat branch satisfies
this integrability condition and admits parallel frames locally, or
globally on a simply connected patch with trivial holonomy. Such
branches include the teleparallel calibration below. A generic non-flat
configuration need not admit a nonzero parallel direction, so this
construction need not yield a charge. This is not a classification of
all possible charges, which would require a characteristic-cohomology
analysis beyond the present work.

\subsection{Reducibility, not the Killing equation}

The condition $D^{\ast}n^a=0$ does not make $n^a$ a reducibility
parameter of the internal redundancy. Preservation of the St\"uckelberg
pair requires $\delta_\epsilon\xi^a=\epsilon^a=0$, even if the
parameter is covariantly constant. Nor is homogeneous parallelism a
specialisation of the generalized-Killing equations for the
coframe--connection configuration.

The connection used in a comparison with metric Killing vectors matters.
For the Levi-Civita connection, $\nabla_\mu\xi_\nu=0$ implies
$\nabla_{(\mu}\xi_{\nu)}=0$, but the converse fails. Schwarzschild
has a timelike Killing vector $\partial_t$, with
$\nabla_{[r}\xi_{t]}=\half f'\neq0$, and no nonzero parallel vector.
It nevertheless admits a homogeneous-parallel internal direction for
the inertial teleparallel pair used in \cref{sec:tegr}. Since
$D^{\ast}$ and $\nabla$ are different connections, homogeneous
parallelism and the metric Killing condition are not ordered in
strength and neither implies the other.

Write an infinitesimal diffeomorphism as a gauge-covariant piece plus a field-dependent
gauge transformation, and add independent homogeneous parameters $(\lambda^{ab},\rho)$. The
covariantised variations are
\begin{align}
 \Delta{b^a}_\mu&=D^{\ast}_\mu\epsilon_\xi^a+\xi^\nu{T^{\ast a}}_{\nu\mu}
 +{\lambda^a}_b\,{b^b}_\mu+\rho\,{b^a}_\mu ,\nonumber\\
 \Delta{A^{ab}}_\mu&=\xi^\nu{R^{ab}}_{\nu\mu}-D_\mu\lambda^{ab} ,\qquad
 \Delta B_\mu=\xi^\nu H_{\nu\mu}-\partial_\mu\rho ,
\label{eq:covvar}
\end{align}
with $\epsilon_\xi^a=\xi^\mu{b^a}_\mu$; these are the standard improved gauge transformations of
Poincar\'e and Weyl gauge theory~\cite{HehlMcCreaMielkeNeeman1995,BlagojevicHehl2013}. Note
the index order in the torsion term: the interior product of a two-form
$T^a=\half{T^{\ast a}}_{\mu\nu}\dd x^\mu\wedge\dd x^\nu$ contributes
$\xi^\nu{T^{\ast a}}_{\nu\mu}\dd x^\mu$, with no additional factor. A
\emph{generalized reducibility parameter} is a triple $(\xi^\mu,\lambda^{ab},\rho)$ for
which the background is invariant,
\begin{equation}
 \Delta{b^a}_\mu=0,\qquad \Delta{A^{ab}}_\mu=0,\qquad \Delta B_\mu=0 ,
\label{eq:reducibility}
\end{equation}
which is the generalized-Killing system for the complete configuration
$({b^a}_\mu,{A^{ab}}_\mu,B_\mu)$. Replacing $\epsilon_\xi^a$ by an
independent $n^a$ and omitting the interior-product and homogeneous terms
would give $D^{\ast}n^a=0$, but would change both the parameter space
and the transformation. In particular, setting $\xi=0$ in the
reducibility system also sets $\epsilon_\xi^a=0$. The two charge
constructions are therefore separate. The coframe equation is
\begin{equation}
 D^{\ast}_\mu\epsilon_\xi^a+\xi^\nu{T^{\ast a}}_{\nu\mu}
 +{\lambda^a}_b\,{b^b}_\mu+\rho\,{b^a}_\mu=0
\label{eq:reducexplicit}
\end{equation}
and will determine the compensators in the following examples.

For Schwarzschild, consider first the static diagonal tetrad in
Weitzenb\"ock gauge, adapted to $\partial_t$ so that
$\Lie_\xi{b^a}_\mu=0$. The triple
$(\xi,\lambda,\rho)=(\partial_t,0,0)$ solves the reducibility system,
with the nonzero derivative of the soldered parameter cancelled by torsion:
\begin{equation}
 D^{\ast}_\mu\epsilon_\xi^a+\xi^\nu{T^{\ast a}}_{\nu\mu}=0 ,
\label{eq:adapted}
\end{equation}
and not by the connection contractions, which vanish identically in that gauge: for
Schwarzschild with $f=1-2M/r$ one has $D^{\ast}_r\epsilon_\xi^{\hat0}=\partial_r\sqrt f$,
exactly cancelled by $\xi^\nu{T^{\ast\hat0}}_{\nu r}$. In a
\emph{non-adapted} frame the compensators are not optional dressing but exactly what turns
a Killing field into a frame symmetry, and \cref{eq:reducexplicit} determines them.
Contracting it with ${h_c}^\mu$ and writing
\begin{equation}
 {M^a}_c\equiv{h_c}^\mu\big(D^{\ast}_\mu\epsilon_\xi^a
 +\xi^\nu{T^{\ast a}}_{\nu\mu}\big) ,\qquad
 \lambda_{ac}+\rho\,\eta_{ac}=-M_{ac} ,
\label{eq:Mdef}
\end{equation}
one has the identity
$M_{ac}=\nabla_c\xi_a+(i_\xi K)_{ac}+(i_\xi B)\,\eta_{ac}$, where $\nabla$ and
$\omega[b]$ are the Levi-Civita connection and spin connection of
$g_{\mu\nu}=\eta_{ab}{b^a}_\mu{b^b}_\nu$ and $K\equiv A-\omega[b]$. Since $\lambda_{ac}$
is antisymmetric, and $K_{ac}$ is too, the sixteen equations separate into $6+1+9$:
\begin{align}
 \lambda_{ac}&=\nabla_{[a}\xi_{c]}-(i_\xi K)_{ac} ,\nonumber\\
 \rho&=-\tfrac14\nabla_\mu\xi^\mu-i_\xi B ,\nonumber\\
 \Lie_\xi g_{\mu\nu}&=\tfrac12\,(\nabla_\rho\xi^\rho)\,g_{\mu\nu} .
\label{eq:compsplit}
\end{align}
The first two equations determine the compensators, and the third is the
conformal Killing condition. The dilation parameter absorbs the trace;
$K$ contributes only to the antisymmetric part and $B$ only to the trace,
so neither occurs in this condition on $\xi$. The bare Kosmann lift
$\lambda^{ab}=\nabla^{[a}\xi^{b]}$ is recovered when $i_\xi K=0$.
Taking $A=\omega[b]$ is sufficient, but not necessary, for this;
a general independent connection requires the correction
$-(i_\xi K)_{ac}$.

Coframe preservation is not yet full reducibility. With $\lambda$ and $\rho$ fixed
by~\cref{eq:compsplit}, $\Delta{A^{ab}}_\mu=0$ and $\Delta B_\mu=0$ are further
conditions on the configuration, and so is the invariance of any retained scalar. The
last is the most telling. For the weight-$(-1)$ compensator of
\cref{sec:multiplet} the covariantised variation is
$\Delta\chi=\xi^\mu D^{\ast}_\mu\chi-\rho\chi$, in which the $i_\xi B$ terms cancel
against those carried by $\rho$, leaving
$\Delta\chi=\xi^\mu\partial_\mu\chi+\tfrac14(\nabla_\mu\xi^\mu)\chi$; preserving a nonzero
\emph{constant} $\chi$ then forces $\nabla_\mu\xi^\mu=0$ and collapses the conformal
Killing condition back to the Killing one. A metric Killing field does satisfy the
conformal Killing condition, and preserves a nonzero constant scalar; what it need not do
is preserve the independent connections $A$ and $B$, or a nonconstant retained scalar. \emph{Asymptotically}, the soldered direction satisfies
$D^{\ast}_\mu\epsilon_\xi^a\to0$ at infinity, which is the regime of the surface integral in
\cref{sec:tegr}.

\subsection{The generalized-reducibility superpotential}
\label{sec:multiplet}

For the generalized-reducibility construction, define the homogeneous
excitations in addition to~\cref{eq:genexc}:
\begin{equation}
 {\mathfrak{H}_{ab}}{}^{\mu\nu}\equiv2\,\frac{\partial\mathcal{L}_{\rm g}}
 {\partial{R^{ab}}_{\mu\nu}},\qquad
 \mathfrak{H}_D{}^{\mu\nu}\equiv2\,\frac{\partial\mathcal{L}_{\rm g}}
 {\partial H_{\mu\nu}} ,
\label{eq:homexc}
\end{equation}
each antisymmetric in $\mu\nu$, so that the boundary term of the first variation is
\begin{equation}
 \Theta^\mu(\delta)={\mathfrak{H}_a}{}^{\mu\nu}\,\delta{b^a}_\nu
 +{\mathfrak{H}_{ab}}{}^{\mu\nu}\,\delta{A^{ab}}_\nu
 +\mathfrak{H}_D{}^{\mu\nu}\,\delta B_\nu ,
\label{eq:bigtheta}
\end{equation}
with unrestricted index sums and no factor of $\half$ on the Lorentz term in this
convention: $\delta\mathcal{L}_{\rm g}=\mathcal{E}\cdot\delta(\text{fields})
+\partial_\mu\Theta^\mu(\delta)$ identically, with ${\mathcal{E}_a}^\mu$,
${\mathcal{E}_{ab}}^\mu$, ${\mathcal{E}_D}^\mu$ and $\mathcal{E}_\chi$ the
Euler-Lagrange densities of $\mathcal{L}_{\rm g}$. We work in the class
\begin{equation}
 \mathcal{L}_{\rm g}=\frac{e}{2}\,\chi^{2}\,\mathbb{T}
 +\frac{e}{4}\,c_R\,R_{ab\mu\nu}R^{ab\mu\nu}
 +\frac{e}{4}\,c_H\,H_{\mu\nu}H^{\mu\nu} ,
\label{eq:multfamily}
\end{equation}
with $\mathbb{T}$ the quadratic torsion family \labelcref{eq:family} and $\chi$ a
compensator scalar of Weyl weight $w=-1$, entering undifferentiated, which makes the
torsion sector strictly scale invariant, the quadratic torsion scalars alone carrying
weight $-2$; $\chi$ is distinct from the matter scalar $\phi$ of
\cref{sec:scalar}. The calibration Lagrangian of \cref{sec:tegr} is the
specialisation in which the scale sector is removed, the compensator is frozen to
$\chi^{2}=\kappa^{-1}$, $c_R=c_H=0$, and the teleparallel rotational branch is taken;
the statements below therefore contain the calibrated model without being confined to
it.

The generator is the full diffeomorphism-Lorentz-dilation multiplet with ordinary
parameters $(\xi^\mu,\omega^{ab},\varrho)$, whose covariantised form
is \cref{eq:covvar} with
\begin{equation}
 \epsilon_\xi^a=\xi^\nu{b^a}_\nu,\qquad
 \lambda^{ab}=\omega^{ab}-\xi^\nu{A^{ab}}_\nu,\qquad
 \rho=\varrho-\xi^\nu B_\nu .
\label{eq:covparams}
\end{equation}
The soldered parameter $\epsilon_\xi^a$ is fixed by the spacetime
generator; it is unrelated to the independent St\"uckelberg parameter,
which leaves the dressed fields inert. In the class~\labelcref{eq:multfamily},
$\Delta\chi=\xi^\mu D^{\ast}_\mu\chi-\rho\chi$, and when $\chi$
is dynamical, exact or asymptotic reducibility also requires
$\Delta\chi=0$ or $\Delta\chi\to0$, respectively. Since $\chi$
occurs without derivatives, it contributes to the dilation Noether
identity but neither to $\Theta^\mu$ nor to the superpotential.
Its Euler derivative $\mathcal{E}_\chi$ consequently has no separate
term in~\cref{eq:multid}. We choose the Noether current
$\mathcal{J}^\mu=\xi^\mu\mathcal{L}_{\rm g}-\Theta^\mu(\delta_{\rm gauge})$,
with the orientation that reproduces~\cref{eq:charge} in the
translational term.

\medskip
\noindent\textbf{Theorem (generalized-reducibility superpotential).}\ \emph{For the
class \labelcref{eq:multfamily} and arbitrary parameter fields, off shell and identically,}
\begin{equation}
 \mathcal{J}^\mu=\partial_\nu\,\mathcal{U}^{\nu\mu}
 +\epsilon_\xi^a{\mathcal{E}_a}^\mu-\lambda^{ab}{\mathcal{E}_{ab}}^\mu
 -\rho\,{\mathcal{E}_D}^\mu ,
\label{eq:multid}
\end{equation}
\begin{equation}
 \boxed{\;\mathcal{U}^{\nu\mu}=\epsilon_\xi^a{\mathfrak{H}_a}{}^{\nu\mu}
 -\lambda^{ab}{\mathfrak{H}_{ab}}{}^{\nu\mu}
 -\rho\,\mathfrak{H}_D{}^{\nu\mu}\;}
\label{eq:multsuper}
\end{equation}
\emph{so that on shell $\mathcal{J}^\mu\eom\partial_\nu\,\mathcal{U}^{\nu\mu}$, and the
multiplet charge is the flux of the superpotential,}
\begin{equation}
 Q[\xi,\lambda,\rho]=\oint_{\partial\Sigma}\mathcal{U}^{i0}\,\dd S_i .
\label{eq:multcharge}
\end{equation}
\noindent The orientation is fixed by the relation between components and forms. With
$\mathrm{vol}_0=\dd x^0\wedge\dd x^1\wedge\dd x^2\wedge\dd x^3$,
$(\dd^3x)_\mu=i_{\partial_\mu}\mathrm{vol}_0$ and
$(\dd^2x)_{\mu\nu}=i_{\partial_\nu}i_{\partial_\mu}\mathrm{vol}_0$, the current and
potential forms are $\mathcal{J}=\mathcal{J}^\mu(\dd^3x)_\mu$ and
$\mathcal{Q}=-\half\,\mathcal{U}^{\mu\nu}(\dd^2x)_{\mu\nu}$, written with the same kernels
as their components and distinguished from them by carrying no indices, as
in~\cref{eq:chargeid}. Then $\dd\mathcal{Q}=\mathcal{J}$ is identically equivalent to
$\mathcal{J}^\mu=\partial_\nu\mathcal{U}^{\nu\mu}$, and on a constant-$x^0$ slice with the
outward boundary orientation $\mathcal{Q}|_{\partial\Sigma}=\mathcal{U}^{i0}\dd S_i$,
which is \cref{eq:multcharge}; the same convention gives \cref{eq:charge} for the
translational term. The presymplectic potential of \cref{eq:bigtheta} enters the same
way, as $\Theta=\Theta^\mu(\dd^3x)_\mu$. No factor of $e$ appears anywhere in
this relation: $\mathcal{U}^{\mu\nu}$ and $\Theta^\mu$ are already coordinate densities
and the $\epsilon$ used here is the alternating \emph{symbol}. Factors of $e$ belong to
the conversion between an ordinary tensor and its density, through the Levi-Civita
\emph{tensor} $e\,\epsilon$, and not to the orientation.

\medskip
\noindent\emph{Derivation.} Substitute \cref{eq:covvar} into \cref{eq:bigtheta}. The
derivative pieces integrate by parts against the antisymmetric excitations,
\begin{align}
 {\mathfrak{H}_a}{}^{\mu\nu}D^{\ast}_\nu\epsilon_\xi^a
 &=-\partial_\nu\big(\epsilon_\xi^a{\mathfrak{H}_a}{}^{\nu\mu}\big)
 +\epsilon_\xi^a\big(\Ds_\nu+T^{\ast}_\nu\big){\mathfrak{H}_a}{}^{\nu\mu},\nonumber\\
 -{\mathfrak{H}_{ab}}{}^{\mu\nu}D_\nu\lambda^{ab}
 &=+\partial_\nu\big(\lambda^{ab}{\mathfrak{H}_{ab}}{}^{\nu\mu}\big)
 -\lambda^{ab}\,\mathcal{D}_\nu{\mathfrak{H}_{ab}}{}^{\nu\mu},
\label{eq:multibp}
\end{align}
and likewise for the $\rho$ term with $\partial_\nu$, each contraction carrying
vanishing total Weyl weight as in \cref{eq:chargeid}, and $\mathcal{D}_\nu$ denoting
the covariant divergence for which the Leibniz rule is exact. This
produces $-\partial_\nu\,\mathcal{U}^{\nu\mu}$ plus parameter contractions of the three
covariant divergences, which are eliminated by the identity forms of the field
equations: $(\Ds_\nu+T^{\ast}_\nu){\mathfrak{H}_a}{}^{\nu\mu}
={(j_{\rm g})_a}^\mu-{\mathcal{E}_a}^\mu$ from \cref{eq:genexc}, and
$\mathcal{D}_\nu{\mathfrak{H}_{ab}}{}^{\nu\mu}
={\mathfrak{H}_{[a}}{}^{\mu\nu}b_{b]\nu}-{\mathcal{E}_{ab}}^\mu$,
$\partial_\nu\mathfrak{H}_D{}^{\nu\mu}
={\mathfrak{H}_a}{}^{\mu\nu}{b^a}_\nu-{\mathcal{E}_D}^\mu$, whose algebraic terms arise
from the $A$- and $B$-dependence of ${T^{\ast a}}_{\mu\nu}$. Those algebraic terms
cancel exactly against the ${\lambda^a}_b\,{b^b}_\nu$ and $\rho\,{b^a}_\nu$
contractions of \cref{eq:covvar} in \cref{eq:bigtheta}, which is why the covariant
regrouping avoids counting the same connection contributions twice. What remains cancels by the index identity
\begin{equation}
 \delta^\mu_\sigma\,\mathcal{L}_{\rm g}
 ={(j_{\rm g})_a}^\mu{b^a}_\sigma
 +{\mathfrak{H}_a}{}^{\mu\nu}{T^{\ast a}}_{\sigma\nu}
 +{\mathfrak{H}_{ab}}{}^{\mu\nu}{R^{ab}}_{\sigma\nu}
 +\mathfrak{H}_D{}^{\mu\nu}H_{\sigma\nu}
\label{eq:indexid}
\end{equation}
contracted with $\xi^\sigma$, which holds because every coordinate index of the
arguments of $\mathcal{L}_{\rm g}$ is displayed in \cref{eq:multfamily} and $\chi$
carries none. We have also verified~\cref{eq:multid} componentwise at generic
exact-rational field and parameter configurations, retaining all five
constants of~\cref{eq:multfamily}.

The superpotential pairs the excitations with the covariant parameters
$(\epsilon_\xi^a,\lambda^{ab},\rho)$ of~\cref{eq:covparams}.
For a pure diffeomorphism, $\omega^{ab}=\varrho=0$, these give
$\lambda^{ab}=-\xi^\nu{A^{ab}}_\nu$ and $\rho=-\xi^\nu B_\nu$.
The connection contractions familiar in tetrad Komar-type potentials
therefore occur as the corresponding terms of this pairing.

For a torsion-only constitutive law the homogeneous excitations vanish
identically, independently of gauge, and the potential reduces to
$Q=\oint\epsilon_\xi^a{\mathfrak{H}_a}{}^{i0}\dd S_i$ for any
$(\lambda,\rho)$. The surface expression used in the TEGR calibration
is thus also obtained from the torsion-only multiplet potential, with
the appropriate parameter. If further conformal symmetries introduce
parameters without independent potentials, there are no additional field
momenta and this three-term expression is unchanged. Alternatively,
an adapted-frame solution of~\cref{eq:reducibility} with
$\lambda=\rho=0$ removes the homogeneous terms for any constitutive
choice in the class. An exactly reducible triple with nonzero
compensators generally retains them.

For asymptotic statements we assume leading asymptotic data rather than bare orders: on
large coordinate spheres $S_r$ in asymptotically Cartesian coordinates, the
configuration is on shell with
\begin{equation}
 \epsilon_\xi^a=\epsilon_{\xi\,\infty}^a+o(1),\qquad
 r^{2}\,{\mathfrak{H}_a}{}^{0i}\to\mathfrak{H}^{(2)\,0i}_a(\theta,\varphi)
\label{eq:falloffdata}
\end{equation}
in an integrable sense on the unit sphere, with
${\mathfrak{H}_{ab}}{}^{0i},\,\mathfrak{H}_D{}^{0i}=O(r^{-2})$, and the generator
translation-type and asymptotically reducible,
\begin{equation}
 \lambda^{ab},\ \rho=o(1),\qquad
 \Theta^i(\delta_{\rm gauge})\,,\ \xi^i\mathcal{L}_{\rm g}=o(r^{-2}) ,
\label{eq:falloffpar}
\end{equation}
the condition on $(\lambda,\rho)$ being a sufficient translation-type asymptotic
condition rather than a canonical asymptotic-symmetry convention.

\medskip
\noindent\textbf{Proposition (asymptotic multiplet charge).}\ \emph{Under
\cref{eq:falloffdata,eq:falloffpar} the limit of \cref{eq:multcharge} over
$S_r$, $r\to\infty$, exists, is conserved, and reduces to its translational term,}
\begin{equation}
 Q[\xi,\lambda,\rho]=\lim_{r\to\infty}
 \oint_{S_r}\epsilon_\xi^a{\mathfrak{H}_a}{}^{i0}\,\dd S_i .
\label{eq:multasy}
\end{equation}

\medskip
\noindent Existence and the reduction follow from the assumed data: the homogeneous
terms contribute $o(1)\times O(r^{-2})\times r^{2}\to0$, while the translational
integrand converges to $\epsilon_{\xi\,\infty}^a\mathfrak{H}^{(2)\,i0}_a$ on the unit sphere.
For conservation, $q^\mu\equiv\partial_\nu\,\mathcal{U}^{\nu\mu}$ is identically
divergence-free, so by Gauss's theorem the change of $Q$ between two slices equals the
flux of $q$, on shell the flux of $\mathcal{J}^\mu$, through the timelike cylinder at
radius $r$, which vanishes as $r\to\infty$ by the last two conditions
of \cref{eq:falloffpar}. Thus the single-translational-term structure of the TEGR
charge extends to translation-type asymptotics across the verified class: the
homogeneous excitations contribute no independent surface term. The value of the
remaining translational charge remains model dependent. Boost- and rotation-type
generators, for which $\lambda^{ab}=O(1)$ or $\xi^\mu$ grows, are not covered by the
hypotheses and are left open, as is the full four-momentum from a boosted solution.

The Noether potential $\mathcal U$ depends on the boundary term chosen
in~\cref{eq:bigtheta}. A Hamiltonian generator instead uses the
variation of the charge together with the presymplectic potential:
in our orientation the form expression is
$\delta\mathcal Q_\eta+i_\xi\Theta$, corresponding to
$k_\eta^{\mu\nu}$ in \cref{sec:admissible}. The relative sign
follows from our definition
$\mathcal J^\mu=\xi^\mu\mathcal L_{\rm g}-\Theta^\mu(\Delta_\eta)$.
Integrability and boundary conditions must be
specified~\cite{Wald1993,IyerWald1994}; agreement of the integrated
Hamiltonian with the Noether flux requires a further charge-matching
argument. The matching established here is the Schwarzschild result in
\cref{sec:tegr}.

A covariant Lagrangian boundary term shifts the Noether potential, but
with the corresponding presymplectic shift its charge variation and
reference-normalised integrable Hamiltonian are unchanged, as proved in
\cref{sec:admissible}. The independent ambiguity
$\Theta\to\Theta+\dd Y(\delta)$ is held fixed in that statement;
it changes the presymplectic current by an exact spacetime term whose
boundary integral need not vanish. Exact reducibility also does not by
itself guarantee a finite bare Noether flux. It gives
$\Theta^\mu(\delta_{\rm gauge})=0$ and
$\mathcal J^\mu=\xi^\mu\mathcal L_{\rm g}$, while leaving
\cref{eq:multid} valid. For example, the pure-tetrad Schwarzschild
representative below requires a reference subtraction.

The theorem is established for~\cref{eq:multfamily}. Its derivation
suggests an extension to other Lagrangians with the field-strength
dependence underlying~\cref{eq:bigtheta}, but we do not establish that
extension here. Away from TEGR the multiplet charge is evaluated only
along the axial deformation for which the Schwarzschild solution
persists. The solution-changing example of
\cref{sec:discriminator} instead uses the parallel-direction charge
in the frozen-compensator, torsion-only theory, rather than the full
dynamical multiplet theorem.

These constructions retain the distinction between local source,
symmetry parameter and surface charge that arises in the linearised
analysis~\cite{HobsonBarkerLasenbyEMT}. The internal redundancy supplies
a strictly internal-$T(4)$-invariant excitation/source system, while the charge requires
additional hypotheses. Neither the parallel-direction construction nor
generalized reducibility has been shown to exhaust all charges, and
neither identifies the source with the external translation class.

\section{Calibration at the TEGR point}
\label{sec:tegr}

The constitutive family fixes a common source-equation structure but
leaves its normalisation and physical interpretation to be checked.
TEGR supplies the comparison with general relativity: its torsion action
is quadratic in first derivatives of the coframe, and its asymptotic
energy is known~\cite{AldrovandiPereira2013,Maluf2013,Moller1961}.
We first identify the excitation and current, then evaluate the
Schwarzschild charge with an inertial connection. The latter is a
calibration against the ADM energy. The spatially flat cosmology and
transverse-traceless wave instead provide consistency checks, through
the Friedmann balance and the averaged Isaacson density. The algebraic
calculations in this section have been checked symbolically. The
non-Einstein example in \cref{sec:discriminator} will give a further
charge assignment, for which independent matching is still required.

\subsection{The excitation and self-current}

At the TEGR point, varying
$\mathcal{L}_{\rm g}=(e/2\kappa)\mathbb T$ with respect to torsion gives
\begin{equation}
 \delta\mathbb{T}=2\,{S_\lambda}^{\mu\nu}\,\delta{T^\lambda}_{\mu\nu} ,
\label{eq:excvar}
\end{equation}
which holds because $S$ is linear in $T$ and $S[T]\!\cdot\!T'$ is symmetric under
$T\leftrightarrow T'$, a symmetry that requires $g^{\mu\nu}$ to be the inverse of
$g_{\mu\nu}$. Hence
${\mathfrak{H}_a}{}^{\mu\nu}=(2e/\kappa)\,{S_a}^{\mu\nu}$: the excitation
\labelcref{eq:excitation} is the standard teleparallel superpotential
\emph{density} in the present normalisation, and \cref{eq:sourceeq} is the teleparallel
field equation.

The off-shell Weitzenb\"ock identity
\begin{equation}
 eR(\{\})=-e\,\mathbb{T}+2\,\partial_\mu(e\,v^\mu)
\label{eq:weitzenbock}
\end{equation}
holds for any coframe and shows that the teleparallel and Einstein-Hilbert actions differ
by a boundary term, the $(\tfrac14,\tfrac12,-1)$ combination being the teleparallel theory
whose field equations reproduce Einstein's. We have checked \cref{eq:weitzenbock} on two
configurations, a spatially flat cosmology with a temporal boundary term and a
static coframe with $R\neq0$ and a spatial boundary term.

Direct evaluation of
${(j_{\rm g})_a}^\mu=\partial\mathcal{L}_{\rm g}/\partial{b^a}_\mu|_\Omega$
at fixed mixed-index ${\Omega^a}_{\mu\nu}$, rather than fixed all-frame
components, also reproduces the standard teleparallel gravitational
current~\cite{deAndradeGuillenPereira2000,Maluf2013}. The same
normalisation holds across its components; this follows from the
coframe variation independently of the identification of the excitation.
\subsection{Schwarzschild energy and the inertial connection}

For the diagonal Schwarzschild
coframe with $f=1-2M/r$ the bare charge diverges,
$I_{\rm bare}(r)=-r\sqrt f\to-(r-M)$, while after subtracting the flat value of the same
coframe
\begin{equation}
 \lim_{r\to\infty}\big(I_{\rm bare}-I_{\rm flat}\big)=M .
\label{eq:admmass}
\end{equation}
The subtraction removes the flat-space contribution and leaves $M$.
For this coframe it is equivalent asymptotically to including the
appropriate inertial spin connection~\cite{KrssakSaridakis2016}.
With $\kappa=8\pi$ no additional constant is fitted. This establishes
the Schwarzschild ADM energy $P^{(0)}=M$; a boosted solution or a
general asymptotic calculation would be required to establish the full
four-momentum.

The current built with the spin connection artificially set to zero is
\emph{not} locally Lorentz covariant: on Minkowski a Cartesian coframe gives vanishing
torsion and vanishing current, while a spherical coframe with the same vanishing connection
gives nonzero torsion and a nonzero bare charge, a pure inertial artefact. These two
choices are not related by a proper covariant gauge transformation. The current built from
the pair (coframe, inertial connection) transforms covariantly, and the normalised charge
is invariant under local Lorentz gauge when coframe, inertial connection and asymptotic
displacement are transformed consistently; its energy component remains observer
dependent, as the time component of any four-vector must.%
\footnote{Explicitly, on Minkowski take
${b^{\hat0}}_\mu\dd x^\mu=\dd t$, ${b^{\hat1}}_\mu\dd x^\mu=\dd r$,
${b^{\hat2}}_\mu\dd x^\mu=r\,\dd\theta$,
${b^{\hat3}}_\mu\dd x^\mu=r\sin\theta\,\dd\phi$ with the inertial connection
$\omega^{\hat1}{}_{\hat2}=-\dd\theta$, $\omega^{\hat1}{}_{\hat3}=-\sin\theta\,\dd\phi$,
$\omega^{\hat2}{}_{\hat3}=-\cos\theta\,\dd\phi$, where indices are moved with
$\omega^a{}_b=\eta_{bc}\omega^{ac}$ and $\omega^{ab}=-\omega^{ba}$, so that for two spatial
indices $\omega^a{}_b=-\omega^b{}_a$. Then
${T^{\ast a}}_{\mu\nu}=2\partial_{[\mu}{b^a}_{\nu]}
+2\omega^a{}_{b[\mu}{b^{|b|}}_{\nu]}=0$ identically, hence zero current and zero charge,
whereas setting $\omega=0$ with the same coframe leaves the nonvanishing anholonomy and a
spurious inertial torsion. The sign hinges on the mixed-index rule just stated.}

The subtraction is not needed
if the charge is evaluated for the covariant pair from the outset. Taking
the same coframe together with the inertial connection of the footnote, the
covariant torsion gives a charge profile finite at every radius outside the
horizon,
\begin{equation}
 I[b,\omega_{\rm inertial}](r)=r\big(1-\sqrt f\,\big),\qquad
 \lim_{r\to\infty}I[b,\omega_{\rm inertial}]=M ,
\label{eq:covcharge}
\end{equation}
with no reference subtraction; on Minkowski the same pair has
${T^{\ast a}}_{\mu\nu}=0$ identically, so the vacuum charge vanishes
identically rather than by cancellation. The reference-subtracted expression~\labelcref{eq:admmass} has the same
asymptotic value. Thus the direct calculation exhibits the equivalence
of the two prescriptions in this example, while the finite-radius
expression~\labelcref{eq:covcharge} is defined by the covariant
coframe--connection pair.

\subsection{Cosmological and radiative checks}

For a spatially flat Robertson--Walker solution in Cartesian comoving
coordinates, ${b^a}_\mu=\mathrm{diag}(1,a,a,a)$, a vanishing spin
connection is the appropriate inertial choice: the torsion vanishes
when $a$ is constant. The pair therefore has a trivial vacuum without
reference subtraction. Evaluation gives
${\mathfrak{H}_{(\hat0)}}{}^{0i}=0$, so the total gravitational and
matter energy within any closed comoving surface vanishes, in agreement
with the teleparallel cosmological result~\cite{Maluf2013}.
The gravitational contribution alone is nonzero. Including the density
factor $e=a^3$, the momentum excitation is
\begin{equation}
 {\mathfrak{H}_{(\hat\imath)}}{}^{0j}=-\frac{2a\dot a}{\kappa}\,\delta_i^{j} ,
\label{eq:frwexc}
\end{equation}
whose undensitised form ${\mathfrak{H}_{(\hat\imath)}}{}^{0j}/e$
has the $\dot a/a^2$ dependence of $S$. Since the density is spatially
constant, its flux through a closed comoving surface vanishes.
The zero energy row of the excitation further requires pointwise
cancellation of the gravitational and matter sources. The explicit
self-current is ${(j_{\rm g})_{(\hat0)}}{}^0=3a\dot a^2/\kappa$,
and the source equation gives
$\rho=3(\dot a/a)^2/\kappa$ for the physical matter density.
Thus the Friedmann constraint appears as the local source balance.
These statements have been verified symbolically and provide a check
on a nonstatic solution without asymptotically flat boundary conditions.

For a linearised transverse-traceless plane wave, take
${b^a}_\mu=\delta^a_\mu+\tfrac12{h^a}_\mu$ in Cartesian coordinates,
with $h_{xx}=-h_{yy}=h_+(t-z)$ and $h_{xy}=h_\times(t-z)$.
The vanishing spin connection again supplies the inertial choice.
The torsion-quadratic self-current first appears at second order,
where its pointwise value in this frame is
\begin{equation}
 {(j_{\rm g})_{(\hat0)}}{}^{0}
 =-\frac{1}{2\kappa}\big(\dot h_+^{\,2}+\dot h_\times^{\,2}\big).
\label{eq:gwdensity}
\end{equation}
With our source-sign convention, the energy density is
$-{(j_{\rm g})_{(\hat0)}}{}^0/e
=(\dot h_+^{\,2}+\dot h_\times^{\,2})/2\kappa$, which is nonnegative
for $\kappa>0$ and positive where the sum of squares is nonzero.
The flow is null, with
${(j_{\rm g})_{(\hat0)}}{}^z={(j_{\rm g})_{(\hat0)}}{}^0$
and zero transverse components. Wavelength averaging gives the Isaacson
energy density and flux. It is this average that provides the
independently established general-relativistic comparison; the
unaveraged expression retains its frame and prescription dependence.

The measured density depends on the observer; a boost changes the
components of the null flux. This should be distinguished from a
simultaneous Lorentz gauge transformation of the coframe and connection.
For exact plane-fronted waves, the teleparallel literature gives a
nonnegative density for static observers~\cite{MalufUlhoa2008} and
vanishing transported energy for a different coframe--connection pair
adapted to the wave~\cite{ObukhovPereiraRubilar2009}. The pairs are
related by a Lorentz transformation of the coframe alone and hence are
not gauge-equivalent configurations. Gauge invariance therefore does
not require their densities to agree. The flux emitted by a bounded
source into an asymptotically flat exterior would provide a further
physical test, left open in \cref{sec:admissible}.

The teleparallel superpotential and current, inertial-connection
prescription~\cite{KrssakSaridakis2016}, reference regularisation and
Schwarzschild mass are established results. Here they follow as the
TEGR specialisation of the internal-source construction, with the same
normalisation also giving the cosmological and wave expressions.
The calibration and consistency checks support that source/charge
interpretation in this sector. They do not add dynamics to the
St\"uckelberg redundancy or establish that an internal $T(4)$ is the
fundamental gravitational gauge group. The internal description remains
an alternative realisation of the coframe coupling.

\section{The charge away from the TEGR point}
\label{sec:discriminator}

We now examine how the current and charge depend on the constitutive
coefficients. The wave and cosmological configurations each determine
one combination of $(\alpha,\beta,\gamma)$. An axial deformation leaves
the Schwarzschild solution and its charge unchanged, whereas a different
exact solution of New General Relativity gives a charge that differs
from the Kepler mass. These evaluations distinguish the common algebraic
structure from its model-dependent values.

\subsection{Wave and cosmological dependence}
Repeating the plane-wave evaluation of \cref{sec:tegr} for the general
family \labelcref{eq:family} gives
\begin{equation}
 {(j_{\rm g})_{(\hat0)}}{}^{0}
 =-\frac{2\alpha+\beta}{2\kappa}\big(\dot h_+^{\,2}+\dot h_\times^{\,2}\big),
\label{eq:gwfamily}
\end{equation}
corresponding to an energy density
$+(2\alpha+\beta)(\dot h_+^{\,2}+\dot h_\times^{\,2})/2\kappa$.
For $\kappa>0$ and $2\alpha+\beta>0$, it is nonnegative and is
positive wherever the sum of squares is nonzero. For example, a linearly
polarised sinusoid has isolated zeros, whereas
$h_+=A\cos u$, $h_\times=A\sin u$ gives the constant sum $A^2$
for the stated circular wave. This concerns the tested
transverse-traceless polarisations and does not establish positivity
for every mode of the theory. We retain the signed current, using an
absolute value only for a stated magnitude.

The plane-wave density thus measures the single combination $2\alpha+\beta$,
independently of $\gamma$: the value \labelcref{eq:gwdensity}, and with it the
Isaacson average, extends across the entire two-parameter subfamily
$2\alpha+\beta=1$. At linear
order the wave torsion is purely tensorial, its vector and axial parts
entering only at $O(h^{2})$. The same evaluation of the cosmological
balance replaces the Friedmann constraint by
\begin{equation}
 \rho=-\tfrac12(2\alpha+\beta+3\gamma)\,\frac{3}{\kappa}
 \Big(\frac{\dot a}{a}\Big)^{2},
\label{eq:frwfamily}
\end{equation}
so the flat cosmology measures $2\alpha+\beta+3\gamma$. The two differ in status: the
cosmological relation is an exact evaluation on the stated ansatz, while the wave relation
holds at quadratic order in the perturbation, which is the order at which the self-current
first appears. Both are statements about the balance law and the current on the stated
configurations; solution persistence away from them is a separate question.

\subsection{The axial deformation}
Within the parity-even quadratic family, deform the constitutive law in the
axial direction,
$(\alpha,\beta,\gamma)\to(\alpha+\tfrac{\lambda}{3},
\beta-\tfrac{2\lambda}{3},\gamma)$, equivalently
$\mathbb{T}\to\mathbb{T}+\lambda A_{\rm ax}$ with
$A_{\rm ax}=T_{[\lambda\mu\nu]}T^{[\lambda\mu\nu]}=\tfrac13(A_1-2A_2)$:
the one-parameter teleparallel family of Hayashi and
Shirafuji~\cite{HayashiShirafuji1979}. On the diagonal Schwarzschild
coframe with its inertial connection the totally antisymmetric torsion part
vanishes identically; the axial excitation, linear in that part, and the
axial self-current, each of whose terms carries at least one factor of it,
then vanish identically as well, both statements verified symbolically. The
configuration therefore remains an exact vacuum solution of the deformed
theory for every $\lambda$, with unchanged excitation and hence unchanged
charge: $Q=M$ across the whole one-parameter family, and
\cref{eq:covcharge} is $\lambda$-independent. The charge therefore remains $M$ away from the TEGR point along a
deformation whose axial torsion is not excited by this solution.
The wave and cosmological evaluations are also insensitive to that
deformation, since it preserves $2\alpha+\beta$ and $\gamma$.
To obtain a different charge in the examples considered here, we next
use a deformation that changes the solution.

\subsection{An exact New General Relativity solution}
Such a deformation is available. Vandeev, Parkhomenko, Semenova and
Matveev~\cite{VandeevEtAl2026} give an exact static spherically symmetric
vacuum solution of New General Relativity, carrying two parameters $M$ and
$k$, on the branch of the same quadratic family given in the notation of
\cref{eq:family} by
\begin{equation}
 \gamma=-\tfrac16\,(2\alpha+\beta)\,(k^{2}+2).
\label{eq:ngrbranch}
\end{equation}
New General Relativity is still a teleparallel theory in the sense used
throughout, with ${R^{ab}}_{\mu\nu}=0$ and the dynamics carried by torsion;
what moves is the constitutive point within \cref{eq:family}, and with it
the solution. It is the equivalence with general relativity, not
teleparallelism, that is given up. In isotropic coordinates
with radial coordinate $\bar r$ the solution is carried by the Cartesian
coframe ${b^{\hat0}}_\mu\dd x^\mu=f\,\dd t$,
${b^{\hat\imath}}_\mu\dd x^\mu=g\,\dd x^i$ in the Weitzenb\"ock gauge, with
\begin{equation}
\begin{aligned}
 f&=\Big(1-\frac{M}{s\bar r}\Big)^{s/2}
    \Big(1+\frac{M}{q\bar r}\Big)^{-q/2},\\
 g&=\Big(1-\frac{M}{s\bar r}\Big)^{1-s/2}
    \Big(1+\frac{M}{q\bar r}\Big)^{1+q/2},
\end{aligned}
\label{eq:ngrsol}
\end{equation}
where $s=4(k-1)/(k-4)$ and $q=-4(k+1)/(k+4)$. We restrict this
parametrisation to $k\notin\{-4,-1,1,4\}$: $s$ and $q$ are singular
at $k=4$ and $k=-4$, respectively, and the family degenerates at
$k=\pm1$. This excludes those parameter values from the displayed
family, without establishing nonexistence of solutions there; their
treatment requires the original equations or an appropriate limit.
We also take the real exterior branch, on which both
$1-M/(s\bar r)$ and $1+M/(q\bar r)$ are positive.

At $k=\pm2$, \cref{eq:ngrbranch} becomes
$\gamma=-(2\alpha+\beta)$. After fixing the overall normalisation,
this is the axial line $2\alpha+\beta=1$, $\gamma=-1$ through TEGR,
and \cref{eq:ngrsol} is the isotropic Schwarzschild solution.
The two families considered here therefore meet at those parameter
values. Since $f^2=1-2M/\bar r+O(M^2/\bar r^2)$, $M$ is the Kepler
mass throughout. The classical tests in Ref.~\cite{VandeevEtAl2026}
constrain departures of $k$ from $2$.

As in the TEGR calculation, the solution is evaluated as a specified
coframe--connection pair. Rotating the coframe locally while retaining
a zero connection is not a gauge transformation and does not in general
preserve the solution. Simultaneously transforming the coframe and
inertial connection is a covariant gauge transformation; we have checked
explicitly that it leaves the charge unchanged.

Evaluating the translational excitation of the general family on this
configuration and taking the surface integral \labelcref{eq:charge} at
infinity gives a closed form,
\begin{equation}
 \frac{Q}{M}=\frac{3(2\alpha+\beta)\,k^{2}}{4(k^{2}-1)} ,
\label{eq:ngrcharge}
\end{equation}
which has been obtained by exact symbolic computation. As a separate
check of the solution and its transcription into our conventions, we
evaluated the source equation at $k=3$ for the energy row and the
independent radial and angular rows, obtaining agreement to between
$50$ and $60$ significant digits at rational sample points. This checks
the coefficient map, excitation, self-current and signs, but is not a
symbolic proof of the field equations throughout the family.

The factor $2\alpha+\beta$ sets the overall action normalisation and
scales the charge linearly. We adopt $2\alpha+\beta=1$, continuing
the TEGR normalisation of \cref{sec:tegr}; the wave calculation
measures the same combination. This is a convention, not a condition
imposed by the dynamics of a general member of the family. With it,
$Q=M$ at $k=\pm2$. At fixed $2\alpha+\beta$ the charge is also
unchanged along the axial direction because the solution has no axial
torsion. Its poles at $k=\pm1$ coincide with the degeneration of the
solution family.

The charge can also be expressed in terms of the post-Newtonian
spatial-curvature parameter. Reading the PPN parameter off the exact isotropic
metric of \cref{eq:ngrsol} gives
$\gamma_{\rm PPN}=(k^{2}+2)/\big(2(k^{2}-1)\big)$, not to be confused with
the constitutive $\gamma$ of \cref{eq:family}, and then, with
$2\alpha+\beta=1$,
\begin{equation}
 \frac{Q}{M}=\frac{1+\gamma_{\rm PPN}}{2},\qquad
 \frac{Q}{M}-1=\frac{\gamma_{\rm PPN}-1}{2}=\frac{4-k^{2}}{4(k^{2}-1)},
\label{eq:ppnidentity}
\end{equation}
exactly on the stated branch. Near $k=2$,
$Q/M=1-\tfrac13(k-2)+\tfrac{13}{36}(k-2)^2+\cdots$.
The fractional difference between the assigned translation charge and
the Kepler mass is thus one half of $\gamma_{\rm PPN}-1$.
This supplies no observational constraint independent of the classical
PPN tests: it expresses the charge assignment in terms of the same
constitutive deviation measured by light propagation. Independent
matching to an asymptotic Hamiltonian charge of this New General
Relativity model remains open and does not follow from the TEGR
normalisation.

The wave and cosmological results determine $2\alpha+\beta$ and
$2\alpha+\beta+3\gamma$, leaving precisely the axial direction
undetermined. The static charge introduces no additional parameter
combination, since $k$ on~\cref{eq:ngrbranch} is fixed by these two.
It provides a distinct static, asymptotic evaluation, and in this
example the deformation changes the solution. All configurations used
here have the stated insensitivity to the axial direction, consistent
with the classical-test behaviour of the axially deformed
family~\cite{HayashiShirafuji1979}. A configuration with nonzero
totally antisymmetric torsion is required to test that remaining
constitutive freedom.

\section{Admissibility tests and open problems}
\label{sec:admissible}

The preceding results suggest the following conditions for a
\emph{physically admissible} translational current. We give them as a
definition for further tests, without claiming that any current has
satisfied the complete set here:
\begin{enumerate}
\item \emph{Gauge covariance}: the excitation/source system transforms
covariantly under the homogeneous factors, and the contracted charge is
invariant when fields, connections and the asymptotic generator are
transformed consistently.
\item \emph{Vacuum triviality}: the current and charge vanish for the
appropriate gauge vacuum, including its inertial connection. This is a
condition on the coframe--connection pair.
\item \emph{Charge matching}: $Q[n]$, or $Q[\xi,\lambda,\rho]$ in the
generalized construction, agrees with an independently defined
Hamiltonian, covariant-phase-space or asymptotic charge of the same
theory. For a non-Einstein theory that charge need not be the ADM value.
\item \emph{Improvement stability}: a Lagrangian boundary term may
change local currents and Noether potentials but leaves the integrable
Hamiltonian charge unchanged at fixed boundary normalisation and
presymplectic representative. The separate exact-form ambiguity
$\mathcal Q\to\mathcal Q+\dd Y$, with $Y$ globally defined on the
integration surface, must also be respected.
\item \emph{Correct weak-field flux}: the current yields the physical
flux of the propagating modes of the theory.
\end{enumerate}

The covariance and vacuum conditions are established for the
teleparallel case above. Charge matching is established for the
Schwarzschild ADM energy, including the axial family that preserves the
solution, but neither for the full four-momentum nor for arbitrary
solutions. The New General Relativity value~\labelcref{eq:ngrcharge}
requires an independent comparison in that theory. The wave calculation
gives the transverse-traceless flux with the Isaacson wavelength average;
emission from a bounded source remains to be established. We now prove
the fourth condition for the normalised charge class.

\emph{Improvement stability.} Let
$\mathcal L'=\mathcal L+\partial_\mu K^\mu$, with $K^\mu$ a local
covariant boundary term and the boundary conditions fixed. We hold the
ordinary generator $\eta=(\xi^\mu,\omega^{ab},\varrho)$ fixed,
including $\delta\xi^\mu=0$. This applies to the Schwarzschild time
translation and the ordinary asymptotic generators considered here.
The covariantised parameters $\lambda^{ab}=\omega^{ab}-\xi\!\cdot\!A^{ab}$
and $\rho=\varrho-\xi\!\cdot\!B$ may still depend on the fields.
Genuinely field-dependent ordinary generators require a corrected
phase-space formula and are not treated.

The field equations are unchanged, and with the same choice of
presymplectic representative,
$\Theta'^\mu=\Theta^\mu+\delta K^\mu$. Thus the presymplectic current
is unchanged because
$\delta_1\delta_2K-\delta_2\delta_1K=0$. This excludes the independent
ambiguity $\Theta\to\Theta+\dd Y(\delta)$, which changes that current
by an exact spacetime term. The Noether potential shifts as
$\mathcal U'^{\mu\nu}=\mathcal U^{\mu\nu}+\xi^\nu K^\mu-\xi^\mu K^\nu$,
so its surface integral can change. The combination
$k_\eta^{\mu\nu}(\delta)=\delta\mathcal U^{\mu\nu}
-\xi^\nu\Theta^\mu(\delta)+\xi^\mu\Theta^\nu(\delta)$
is nevertheless unchanged. To fix its divergence and signs, define
\begin{equation}
 W^\mu(\delta,\Delta_\eta)=\delta\big[\Theta^\mu(\Delta_\eta)\big]
 -\Delta_\eta\big[\Theta^\mu(\delta)\big] ,
\label{eq:presymp}
\end{equation}
with the ordinary generator $\eta=(\xi^\mu,\omega^{ab},\varrho)$ held fixed, so that
$\Delta_\eta$ commutes with $\delta$. Varying
$\mathcal{J}^\mu=\xi^\mu\mathcal{L}_{\rm g}-\Theta^\mu(\Delta_\eta)$, and using
$\delta\mathcal{L}_{\rm g}=\mathcal{E}_I\delta\phi^I+\partial_\nu\Theta^\nu$ together with
the transformation law of the internally invariant vector density $\Theta^\mu$, gives
\begin{equation}
 \delta\mathcal{J}^\mu=\xi^\mu\mathcal{E}_I\delta\phi^I
 +\partial_\nu\big(\xi^\mu\Theta^\nu-\xi^\nu\Theta^\mu\big)-W^\mu .
\label{eq:deltaJ}
\end{equation}
Varying \cref{eq:multid} in the form
$\mathcal{J}^\mu=\partial_\nu\mathcal{U}^{\nu\mu}+C_\eta^\mu$, with $C_\eta^\mu$ the Euler
combination there, and subtracting, one obtains identically
\begin{equation}
 \partial_\nu k_\eta^{\nu\mu}
 =-W^\mu+\xi^\mu\mathcal{E}_I\delta\phi^I-\delta C_\eta^\mu .
\label{eq:kdiv}
\end{equation}
For a solution and a variation satisfying the linearised field equations,
this gives $\partial_\nu k_\eta^{\nu\mu}=-W^\mu(\delta,\Delta_\eta)$.
The divergence acts on the first index; its sign follows from the
definition~\labelcref{eq:presymp}. If the generator preserves the complete
background, including any retained scalar, its tangent vector vanishes
and hence $W^\mu=0$. The charge-variation form is then closed, without
requiring the perturbation $\delta$ itself to preserve that symmetry.

For fixed ordinary parameters, the variation of the shift
$\xi^\nu K^\mu-\xi^\mu K^\nu$ cancels the additional
$\xi^\nu\delta K^\mu-\xi^\mu\delta K^\nu$ from $\Theta'$.
One must retain the variations of the covariantised parameters,
\begin{equation}
 \delta\epsilon_\xi^a=\xi^\nu\delta{b^a}_\nu,\qquad
 \delta\lambda^{ab}=-\xi^\nu\delta{A^{ab}}_\nu,\qquad
 \delta\rho=-\xi^\nu\delta B_\nu ,
\label{eq:covparamvar}
\end{equation}
inside $\delta\mathcal U^{\mu\nu}$; the $\xi\Theta$ terms combine
with them. The notation $k_\eta^{\mu\nu}$ reverses the displayed
index order of $\mathcal U^{\nu\mu}$ in~\cref{eq:multid}, without
introducing another sign. Whenever $k_\eta$ is integrable on the chosen
phase space, the Hamiltonians for the two Lagrangians can differ only
by a field-independent constant. Assigning the same value to a common
reference removes this difference; here we use
$H_\xi[\mathrm{Minkowski}]=0$.

Separately, $\mathcal Q\to\mathcal Q+\dd Y$ changes no integral
over a closed surface if $Y$ is globally defined there, by Stokes'
theorem. This requires no asymptotic limit. Thus the fixed multiplicative
normalisation $\kappa=8\pi$ and common additive normalisation give
the same integrable Hamiltonian for every representative related by
the stated boundary terms. The Schwarzschild value $M$ calibrates this
normalisation; the proof of stability does not depend on defining it
through that value.

In the scale-fixed, first-order, parity-even quadratic torsion sector,
consider a boundary vector local and covariant in the coframe and torsion,
without torsion derivatives, matter fields or background vectors, and
with the required dimension and weight. It is linear in torsion, whose
only parity-even vector contraction is the trace. Hence
$K^\mu=c\,e\,v^\mu$, up to normalisation; the axial contraction is
parity odd. This is the boundary term in the Weitzenb\"ock
identity~\labelcref{eq:weitzenbock}, so the stability statement includes
the relation between the teleparallel and Einstein--Hilbert actions.
Allowing the compensator $\chi$ changes the available expressions and
would require a broader classification.

For the Schwarzschild covariant pair, this boundary term shifts the
asymptotic Noether flux of $\xi=\partial_t$ to $M(1+4\pi c)$, while
the phase-space variation gives $\dd H/\dd M=1$ at every radius.
The reference-normalised Hamiltonian is therefore $H=M$ for every
$c$ and $r$. It should be distinguished from either finite-radius
Noether flux. The parallel unit direction
$n^{\hat a}=\delta^{\hat a}_{\hat0}$ gives
$Q[n](r)=r(1-\sqrt f)$, whereas the soldered parameter of
$\partial_t$ is $\epsilon_\xi^{\hat0}={b^{\hat0}}_t=\sqrt f$ and
gives $Q[\partial_t](r)=r\sqrt f(1-\sqrt f)$ at $c=0$.
These two fluxes agree only asymptotically.

Invariance of the presymplectic current also implies that the stated
Lagrangian boundary terms leave a radiative flux defined by that
structure unchanged. This relates improvement stability to the flux
condition, but does not establish emission from a bounded source.
The explicit statements above have been checked by exact symbolic
calculation. None requires the pointwise densities in
\cref{sec:tegr} to be representative independent.

\section{Discussion}
\label{sec:discussion}

The results distinguish several questions that arise in discussions of
local gravitational energy. Teleparallel constructions provide local
self-currents and surface energies~\cite{Moller1961,Maluf2013,AldrovandiPereira2013},
whereas the metric results of Ref.~\cite{HobsonBarkerLasenbyEMT}
exclude strict representatives of specified linearised external
translation classes. The present construction concerns the coframe
source of an independent-connection theory and does not contradict that
exclusion. In particular, introducing an internal redundancy changes
neither the existing homogeneous transformation laws nor the underlying
dynamics.

With the gravitational/matter split and boundary representative fixed,
the coframe equation identifies a local gravitational self-current.
Its frame-indexed form transforms homogeneously under Lorentz and
dilation transformations; the soldered density~\labelcref{eq:soldered}
is invariant under the internal gauge factors. This makes precise one
sense in which a local gravitational density is available. To identify
it as a preferred physical energy observable would require further
information about conservation, representative selection and a
relational, diffeomorphism-invariant interpretation; we disagree with
any reading in which the density is an observer-independent physical
energy density.

The teleparallel calibration establishes more than the formal
existence of a source. It reproduces the known superpotential, current
and Schwarzschild ADM energy for a covariant coframe--connection pair,
and gives the standard cosmological balance and averaged wave density.
The structural source equation and self-current extend throughout the
parity-even constitutive family. The generalized-reducibility potential
also allows homogeneous-sector contributions outside the torsion-only
case. These extensions concern the form and conditional charges of the
source, rather than a universal energy observable independent of its
prescription.

The charge tests have correspondingly different scopes. Independent
matching has been demonstrated for Schwarzschild at TEGR and along the
axial deformation that preserves that solution. The New General
Relativity calculation gives a definite assignment related to the
Kepler mass by~\cref{eq:ppnidentity}, but requires comparison with an
independently defined charge in that model. The improvement theorem
establishes stability of an integrable, reference-normalised Hamiltonian
under the specified Lagrangian boundary terms; it does not select a
unique local current or remove the independent presymplectic ambiguity.
Likewise, the sufficient conditions for charges do not establish their
existence on every configuration.

The distinction between soldering and gauge fixing is particularly
important here. Earlier work on linearised metric gravity obtained
local densities by imposing gauge
conditions~\cite{ButcherHobsonLasenby2010,ButcherLasenbyHobson2012,ButcherHobsonLasenby2012}.
The source construction instead defines the self-current before such a
choice, and contracts its frame index with a dynamical field.

Soldering introduces no external direction and retains the internal
symmetries. Setting the spin connection to zero while treating a local
coframe rotation as a change of the coframe alone gives the pure-tetrad
expression, whose local Lorentz covariance fails as described in
\cref{sec:tegr}. A consistent transformation of the complete pair
preserves covariance, while a change of observer can still change
measured components. Internal gauge invariance, coordinate covariance
and observer independence are therefore distinct requirements.

Ordinary conservation of the soldered gravitational density also does
not follow from~\cref{eq:balance}. That equation concerns the total
frame-indexed source; forming $\widehat{\jmath}$ both removes the
matter contribution and introduces a coframe factor. Differentiation
gives
\begin{equation}
\begin{split}
 \partial_\mu\widehat{\jmath}_\nu{}^\mu
 ={}&{b^a}_\nu\big[(\Ds_\mu+T^{\ast}_\mu){J_a}^\mu
 -(\Ds_\mu+T^{\ast}_\mu){(t_{\rm m})_a}^\mu\big]\\
 &+\big(\Ds_\mu{b^a}_\nu\big)\,{(j_{\rm g})_a}^\mu ,
\end{split}
\label{eq:soldereddiv}
\end{equation}
where the matter exchange and the derivative of the coframe occur
separately. Here $\Ds_\mu{b^a}_\nu$ acts only on the internal index
and Weyl weight. It is not the full tetrad-postulate derivative and
need not vanish; its antisymmetric part is the torsion
in~\cref{eq:covtorsion}. Even on a Lorentz- and scale-flat branch,
where the total frame-indexed source is covariantly conserved, this
last term remains. Thus flat homogeneous connections do not suffice
to establish ordinary conservation of the soldered self-current.
The terms may vanish or cancel on particular solutions, including
configurations with nonzero curvature; the identity is not a general
nonconservation theorem.

This separates the result from the Maxwell benchmark. The Maxwell
stress tensor is internally gauge invariant, conserved on shell and a
representative of the external translation Noether class. For the
soldered gravitational self-current we have established the first
property, but not the other two in general. Nor does the source equation
select an action split or boundary representative. Ordinary Yang-Mills
theory has no canonical coframe to contract its internal charge-current
index into a coordinate index, so that particular comparison cannot
establish an obstruction to the gravitational construction.

The torsion-free reduction supplies a further, limited connection with
the metric analysis. \Cref{app:torsionfree} shows that a
conformal curvature term reduces to its metric Weyl-squared counterpart.
The isolated reduced metric theory then falls within the linearised
result~\cite{HobsonBarkerLasenbyEMT}, but no descent of the full reduced
WGT field equations or current-equivalence classes is established.
Similarly, generalized canonical constructions retaining boundary
terms in metric gravity~\cite{Kenmoku2026} concern the external
coordinate-translation representative, rather than identifying it
with the internal source considered here. The covariance of the
frame-indexed source also suggests a comparison with the antisymmetric
parameter in the metric compensator calculation, but supplies no
identification of those mechanisms or classes. It follows the general
indexed-current transformation laws of the covariant
formalism~\cite{HobsonLasenbyBarker2024}.

The internal interpretation therefore makes the invariance of the
existing coframe source manifest and provides the source equation,
constitutive dependence and conditional charge constructions studied
above. It gives no new variational current merely by enlarging the
redundant field description. Further work is required on ordinary
conservation of the soldered density, full four-momentum, charges of
generic solutions and independent matching in non-Einstein models.
The eWGT extension also requires a new analysis: its dilation terms
modify the translational field strength and derivative themselves~\cite{eWGTpaper},
so its kinematics cannot be obtained merely by relabelling the present
expressions.

\section{Conclusions}
\label{sec:conclusions}

We have examined the coframe source and its charges when translations
are introduced as an internal St\"uckelberg $T(4)$ redundancy in WGT.
The dressed coframe and its covariant field strengths are invariant under
the introduced internal $T(4)$, and variation of the translational potential gives the existing coframe
Euler derivative. No flatness condition is required, and the redundancy
adds no dynamics. The coframe equation takes an excitation/source form,
with an ordinarily conserved total source in the translations-only
sector and a covariant balance law in the full theory. The explicit
self-current has a common algebraic form throughout the parity-even
quadratic torsion family, with values determined by the constitutive
coefficients.

Contracting the self-current with the coframe gives a local density
invariant under the internal gauge factors, relative to the specified
action split and boundary representative. This does not establish its
ordinary conservation, select a preferred density or identify it with
the external translation Noether class. The separate roles of local
source and charge remain necessary: we derive a translational surface
charge using a covariantly constant internal direction, whose existence
is constrained by homogeneous holonomy, and a three-sector Noether
potential for generalized reducibility of the complete scale-invariant
quadratic configuration. Neither construction is proved necessary or
obtained by restricting the other. Conservation of the asymptotic
multiplet flux uses the stated falloff and flux conditions; on an
exact-symmetry background the charge-variation form is closed for
linearised solutions.

At TEGR the excitation and self-current reproduce the teleparallel
expressions, and the covariant Schwarzschild pair gives the ADM energy.
The spatially flat cosmology has vanishing total comoving energy through
the pointwise Friedmann balance, while the transverse-traceless wave
has null flow and the Isaacson density after wavelength averaging.
Away from TEGR, these configurations determine the combinations
$2\alpha+\beta$ and $2\alpha+\beta+3\gamma$. The Schwarzschild
charge is unchanged by an axial deformation, whereas an exact New
General Relativity solution gives
$Q/M=(1+\gamma_{\rm PPN})/2$ with the adopted normalisation.
This is a definite charge assignment whose independent Hamiltonian
matching remains open; it supplies no additional PPN constraint.

For fixed ordinary generators, boundary conditions and presymplectic
representative, covariant Lagrangian boundary terms leave the charge
variation unchanged. An integrable Hamiltonian consequently has the
same value after fixing a common reference normalisation, although the
local current and bare Noether potential can change. This establishes
one of the admissibility conditions proposed here. Further tests include
the full ADM four-momentum from a boosted solution, radiation from a
bounded source, independent charge matching in non-Einstein models,
and configurations with nonzero axial torsion. The extension to eWGT,
whose torsion-trace modification changes the translational sector,
requires a separate treatment and is considered in a forthcoming
companion paper.

\begin{acknowledgments}
W.~B. is grateful for the support of Girton College, Cambridge, Marie
Sk\l{}odowska-Curie Actions, and the hospitality of the Helsinki Institute of
Physics. Co-funded by the European Union (Physics for Future -- Grant Agreement
No.\ 101081515). Views and opinions expressed are, however, those of the authors only
and do not necessarily reflect those of the European Union or European Research
Executive Agency. Neither the European Union nor the granting authority can be held
responsible for them.
The authors used Claude (Anthropic; Fable 5.1) and ChatGPT (OpenAI;
GPT-6 Astra) for
checking analytical calculations, writing code for symbolic and numerical
computations, literature search, and drafting suggestions; all results were
independently verified, and the authors take full responsibility for the content.
\end{acknowledgments}

\appendix
\crefalias{section}{appendix}

\section{The torsion-free WGT branch}
\label{app:torsionfree}

We consider the branch ${T^{\ast a}}_{\mu\nu}=0$ and determine the
reduced conformal-curvature action. Its isolated metric sector can be
compared with the metric localisation theorem, but that comparison does
not establish a relation between the current classes of the full
reduced WGT and the isolated theory. The constraint is imposed here to
examine this branch, without asserting that it is physically preferred.

\emph{The reduced connection is a Weyl connection.} Because the covariant
derivative~\labelcref{eq:covtorsion} of the weight-one coframe carries the dilation gauge
field, solving ${T^{\ast a}}_{\mu\nu}=0$ for the rotational connection returns
not the Ricci rotation coefficients alone but the \emph{Weyl connection}
\begin{equation}
 A_{ab\mu}=\omega_{ab\mu}[b]
 +\big(\eta_{ac}\mathcal{B}_b-\eta_{bc}\mathcal{B}_a\big){b^c}_\mu ,
 \qquad \mathcal{B}_a={h_a}^\mu B_\mu ,
\label{eq:weylconnapp}
\end{equation}
with $\omega_{ab\mu}[b]$ the Ricci rotation coefficients of the coframe: the connection
is solved as $A=A[b,B]$, not $A=A[b]$.

\emph{Curvature decomposition.} Substituting~\cref{eq:weylconnapp}, the rotational
curvature differs from the metric Riemann tensor by wholly trace-generated terms,
\begin{equation}
 \mathcal{R}_{abcd}\big[A[b,B]\big]=R_{abcd}[g]+(\eta\owedge\mathcal{S})_{abcd},
\label{eq:reduceapp}
\end{equation}
with
\begin{equation}
 \mathcal{S}_{ad}=-\nabla_d\mathcal{B}_a+\mathcal{B}_a\mathcal{B}_d
 -\half\mathcal{B}^2\eta_{ad},
\label{eq:reduceSapp}
\end{equation}
where $\nabla$ is the metric (Levi-Civita) covariant derivative and
\begin{equation}
 (\eta\owedge\mathcal{S})_{abcd}
 =\eta_{ac}\mathcal{S}_{bd}+\eta_{bd}\mathcal{S}_{ac}-\eta_{ad}\mathcal{S}_{bc}-\eta_{bc}\mathcal{S}_{ad}
\label{eq:kndefapp}
\end{equation}
is a Kulkarni--Nomizu-type combination, here \emph{generalized}: $\mathcal{S}_{ad}$ is not
symmetric, its antisymmetric part being the dilation field strength,
$\mathcal{S}_{[ad]}=\nabla_{[a}\mathcal{B}_{d]}=\half\mathcal{H}_{ad}$, with
$\mathcal{H}_{ad}={h_a}^\mu{h_d}^\nu H_{\mu\nu}$. The
decomposition~\labelcref{eq:weylconnapp}--\labelcref{eq:kndefapp} has been verified symbolically
at generic exact-rational field configurations.

\emph{Consequences.} Every term of~\cref{eq:reduceapp} beyond $R_{abcd}[g]$ carries an
explicit $\eta$, so the difference is entirely trace generated and its totally
trace-free projection vanishes:
\begin{equation}
 \mathrm{Weyl}\big(\mathcal{R}[A[b,B]]\big)=C[g].
\label{eq:weylprojapp}
\end{equation}
The symmetric part $\mathcal{S}_{(ad)}$ is the ordinary Weyl-connection trace deformation and
preserves the Riemannian index symmetries; the antisymmetric part does not, so the
pair-exchange symmetry and the first Bianchi identity, absent a priori from a gauge
curvature, are \emph{not} recovered on the branch, their violations being governed
precisely by $\mathcal{H}_{ad}$ and vanishing only where the dilation field strength
does. At the level of the action, \cref{eq:weylprojapp} gives
\begin{equation}
 \mathrm{Weyl}(\mathcal{R})_{abcd}\,\mathrm{Weyl}(\mathcal{R})^{abcd}
 \;\longrightarrow\;C_{abcd}[g]\,C^{abcd}[g]\,:
\end{equation}
a conformal (Weyl-squared) curvature term reduces on this branch to its metric
counterpart, whose linearisation about flat space admits no strict local
gauge-invariant rank-two representative of the translation Noether
class~\cite{HobsonBarkerLasenbyWS,HobsonBarkerLasenbyEMT}. This consequence applies to the isolated reduced metric theory about
flat space. Substitution of the torsion-free connection into the action
does not establish a descent of the full reduced WGT field equations,
current equivalences or improvement classes, which still involve the
surviving fields and their Euler derivatives. No nonlinear or
full-theory localisation result follows from the action reduction alone.

Only the totally trace-free curvature becomes metric. Ricci-type
invariants generally retain $B_\mu$ through $\mathcal S_{ad}$, and
the dilation sector retains its first-order field strength $H_{\mu\nu}$.
Away from the torsion-free branch the rotational connection is independent
and this reduction does not apply. The teleparallel branch of
\cref{sec:internal} instead retains torsion and the first-order
excitation/source construction of the main text.


\begin{thebibliography}{99}


\bibitem{HobsonBarkerLasenbyFP}
M.P.~Hobson, W.E.V.~Barker and A.N.~Lasenby,
Phys.~Rev.~D {\bf 114}, 024033 (2026).

\bibitem{HobsonBarkerLasenbyWS}
M.P.~Hobson, W.E.V.~Barker and A.N.~Lasenby, ``Bessel-Hagen currents for
linearised Weyl-squared gravity'', arXiv:2608.02912.

\bibitem{HobsonBarkerLasenbyEMT}
M.P.~Hobson, W.E.V.~Barker and A.N.~Lasenby, ``Bessel-Hagen currents,
reducibility charges, and local gravitational energy'', companion paper (2026).

\bibitem{HobsonLasenbyBarker2024}
M.P.~Hobson, A.N.~Lasenby and W.E.V.~Barker,
Phys.~Rev.~D {\bf 109}, 024022 (2024).

\bibitem{CGTpaper}
M.P.~Hobson and A.N.~Lasenby,
Phys.~Rev.~D {\bf 103}, 104042 (2021).

\bibitem{eWGTpaper}
A.N.~Lasenby and M.P.~Hobson,
J.~Math.~Phys.~{\bf 57}, 092505 (2016).

\bibitem{Kibble1961}
T.W.B.~Kibble, J.~Math.~Phys.~{\bf 2}, 212 (1961).

\bibitem{BlagojevicHehl2013}
M.~Blagojevi\'c and F.W.~Hehl (eds.), \textit{Gauge Theories of Gravitation: A Reader with
Commentaries} (Imperial College Press, London, 2013).

\bibitem{HehlMcCreaMielkeNeeman1995}
F.W.~Hehl, J.D.~McCrea, E.W.~Mielke and Y.~Ne'eman,
Phys.~Rept.~{\bf 258}, 1 (1995).

\bibitem{ItinHehlObukhov2017}
Y.~Itin, F.W.~Hehl and Y.N.~Obukhov,
Phys.~Rev.~D {\bf 95}, 084020 (2017).

\bibitem{KoivistoZlosnik2022}
T.~S. Koivisto and T.~Z\l{}o\'snik,
``Paths to gravitation via the gauging of parameterized field theories,''
arXiv:2212.04562.

\bibitem{KoivistoZlosnik2024}
T.~S. Koivisto and T.~Z\l{}o\'snik,
``Gravity from pre-geometry,''
arXiv:2409.02200.

\bibitem{Obukhov2018}
Y.N.~Obukhov, Int.~J.~Geom.~Methods Mod.~Phys.~{\bf 15}, 1840005 (2018).

\bibitem{AldrovandiPereira2013}
\interlinepenalty=10000
R.~Aldrovandi and J.G.~Pereira, \textit{Teleparallel Gravity: An Introduction}
(Springer, Dordrecht, 2013).

\bibitem{KrssakSaridakis2016}
M.~Kr\v{s}\v{s}\'ak and E.N.~Saridakis,
Class.~Quantum Grav.~{\bf 33}, 115009 (2016).

\bibitem{deAndradeGuillenPereira2000}
V.C.~de Andrade, L.C.T.~Guillen and J.G.~Pereira,
Phys.~Rev.~Lett.~{\bf 84}, 4533 (2000).

\bibitem{Wald1993}
R.M.~Wald, Phys.~Rev.~D {\bf 48}, R3427 (1993).

\bibitem{IyerWald1994}
V.~Iyer and R.M.~Wald, Phys.~Rev.~D {\bf 50}, 846 (1994).

\bibitem{Maluf2013}
J.W.~Maluf, Ann.~Phys.~(Berlin) {\bf 525}, 339 (2013).

\bibitem{Moller1961}
C.~M\o ller, Ann.~Phys.~(N.Y.)~{\bf 12}, 118 (1961).

\bibitem{MalufUlhoa2008}
J.~W. Maluf and S.~C. Ulhoa,
``The energy-momentum of plane-fronted gravitational waves in the
teleparallel equivalent of GR,''
Phys.\ Rev.\ D \textbf{78}, 047502 (2008);
Erratum: Phys.\ Rev.\ D \textbf{78}, 069901 (2008).

\bibitem{ObukhovPereiraRubilar2009}
Y.~N. Obukhov, J.~G. Pereira, and G.~F. Rubilar,
``On the energy transported by exact plane gravitational-wave solutions,''
Class.\ Quantum Grav.\ \textbf{26}, 215014 (2009).

\bibitem{HayashiShirafuji1979}
K.~Hayashi and T.~Shirafuji,
``New general relativity,''
Phys.\ Rev.\ D \textbf{19}, 3524 (1979);
Addendum: Phys.\ Rev.\ D \textbf{24}, 3312 (1981).

\bibitem{VandeevEtAl2026}
V.~P. Vandeev, V.~A. Parkhomenko, A.~N. Semenova, and M.~A. Matveev,
``Classical tests of New General Relativity,''
arXiv:2605.05500 [gr-qc].

\bibitem{ButcherHobsonLasenby2010}
L.M.~Butcher, M.P.~Hobson and A.N.~Lasenby,
Phys.~Rev.~D {\bf 82}, 104040 (2010).

\bibitem{ButcherLasenbyHobson2012}
L.M.~Butcher, A.N.~Lasenby and M.P.~Hobson,
Phys.~Rev.~D {\bf 86}, 084012 (2012).

\bibitem{ButcherHobsonLasenby2012}
L.M.~Butcher, M.P.~Hobson and A.N.~Lasenby,
Phys.~Rev.~D {\bf 86}, 084013 (2012).

\bibitem{Kenmoku2026}
M.~Kenmoku, ``Covariant energy momentum tensor in general relativity by
generalized canonical method'', arXiv:2608.18459 [gr-qc].

\end{thebibliography}
\end{document}